\documentclass[11pt]{article}

\usepackage[utf8]{inputenc}
\usepackage{float}
\usepackage{setspace}
\usepackage[margin=1in]{geometry}
\usepackage{caption}
\usepackage{subcaption}
\usepackage{hyperref}
\usepackage[open,openlevel=1]{bookmark}

\usepackage{authblk}
\usepackage{natbib}
\usepackage{xcolor}
\usepackage{amsmath}
\usepackage{amssymb}
\usepackage{graphicx} 
\usepackage{adjustbox}
\usepackage[normalem]{ulem}
\useunder{\uline}{\ul}{}
\usepackage{bm}

\usepackage{accents}
\renewcommand{\vec}[1]{\underaccent{\tilde}{#1}}

\begin{document}

\title{Multidimensional Poverty Mapping for Small Areas}

\author[1]{Soumojit Das\footnote{Corresponding Author. Email: soumojit.das@wsu.edu}}
\author[2]{Dilshanie Deepawansa}
\author[3]{Partha Lahiri}

\affil[1]{Washington State University, Pullman, WA, USA}
\affil[2]{Department of Census and Statistics, Sri Lanka}
\affil[3]{University of Maryland, College Park, MD, USA}

\date{\vspace{-5ex}}

\maketitle

\begin{abstract}
Many countries measure poverty based only on income or consumption. However, there is a growing awareness of measuring poverty through multiple dimensions that captures a more reasonable status of poverty. Estimating poverty measure(s) for small geographical areas, commonly referred to as poverty mapping, is challenging due to small or no sample for the small areas.  While there is a huge literature available on unidimensional poverty mapping, only a limited effort has been made to address special challenges that arise only in the multidimensional poverty mapping. For example, in multidimensional poverty mapping, a new problem arises involving estimation of relative contributions of different dimensions to overall poverty for small areas. This problem has been grossly ignored in the small area estimation (SAE) literature. We address this issue using a multivariate hierarchical model implemented via a Bayesian method.  Moreover, we demonstrate how a multidimensional poverty composite measure can be estimated for small areas. In this paper, we demonstrate our proposed methodology using a survey data specially designed by one of us for multidimensional poverty mapping.   This paper adds a new direction to poverty mapping literature.

\end{abstract}


\section*{Statement of Significance}

This paper addresses a critical gap in poverty measurement by developing novel statistical methods for estimating multidimensional poverty at small geographic areas. While existing literature focuses on unidimensional poverty mapping, we introduce multivariate hierarchical models that not only estimate poverty rates but also quantify the relative contributions of different poverty dimensions (e. g., material deprivation, social deprivation, human capital, etc.) to overall poverty status. Our hierarchical Bayesian approach handles the challenge of limited sample sizes in small areas by incorporating auxiliary information from census data and properly accounting for sampling variance uncertainty. The methodology enables policymakers to identify which dimensions drive poverty in specific regions, facilitating targeted interventions aligned with United Nation's Sustainable Development Goal 1. We demonstrate the practical application using survey data from Sri Lanka's Uva province, providing reliable estimates even for areas with minimal sample coverage, and produce detailed poverty maps that guide evidence-based policy decisions.

\section{Introduction}

The United Nations (UN) established the Sustainable Development Goals (SDGs) in 2015, which comprise 17 goals and 169 targets. These goals are part of the 2030 Agenda for Sustainable Development and represent a global effort to encourage all countries and stakeholders to collaborate in order to eliminate poverty, preserve the environment, and promote peace and prosperity for all people by the year 2030. These 17 SDGs have more than 210 indicators. These indicators are used to measure progress towards achieving the goals and cover a wide range of topics, including poverty, health, education, gender equality, and sustainable development. Each goal has a specific set of indicators, and some indicators are shared across multiple goals. Reliable disaggregated data is critical for achieving Sustainable Development Goals because the goals aim to be inclusive and comprehensive. Disaggregating data by factors such as age, gender, income, and location helps policymakers identify specific needs and challenges faced by different groups and design tailored policies and programs to address them. This approach ensures that all segments of society are accounted for and no one is left behind in the pursuit of the SDGs. 

The very first of these 17 goals is `No Poverty: End poverty in all its forms everywhere.' The primary indicators for this goal include measuring the proportion of the population living below the international poverty line, the proportion of the population covered by social protection systems, and the proportion of the population living in households with access to basic services. These indicators help measure progress towards the goal and inform policy development. There are multiple approaches to conceptualizing poverty: unidimensional poverty and multidimensional poverty. Unidimensional poverty measures are based on a single dimension, usually income or consumption. Such monetary approach is the earliest attempt at defining poverty by assessing individuals or household welfare in terms of income or consumption. The multidimensional approach considers the set of different disadvantages experienced by poor people simultaneously, such as deprivation in health, education, nutrition, and lack of clean water, etc. 

Sen's Capability approach (\cite{sen1980equality}) provides a broad normative framework for assessing individual well-being and the nature of poverty. This laid the philosophical foundations for the multidimensional poverty measures that followed. The {\em Fuzzy Set} method (see \cite{cerioli1990fuzzy}) and {\em Counting} method (see \cite{alkire2015multidimensional}) are predominantly used to measure multidimensional poverty.  Recently, \cite{deepawansa2022innovative} proposed a {\em Synthesis} method, which is a well-organized amalgamation of these two techniques.  Multidimensional poverty measures have become feasible due to the availability of household survey data. Multidimensional measures of poverty allow for the quantification of poverty from different dimensions by incorporating multiple indicators of poverty. This enables a more comprehensive understanding of poverty, including its multiple dimensions and how they interact, which, in turn, can inform more effective poverty reduction strategies. These measures can also identify the most disfavored individuals or groups, enabling targeted interventions to address their specific needs. Such multidimensional measures of poverty offer more than just an overall measure that combine various dimensions. They can also help quantify the impact of individual dimensions, which can help identify areas of concern and allocate resources accordingly to address issues associated with each dimension. The multidimensional poverty index (MPI) captures various deprivations experienced simultaneously by the poor. It is crucial to understand the contribution of each dimension to total poverty for high-impact policy design and effective targeting. For instance, while two geographical areas might have similar MPI rates, the dimensions contributing to their poverty could differ significantly, necessitating different policy and budgetary responses. Understanding these contributions helps in identifying policy priorities at the lowest administrative levels, aligning with the SDG goal 1 of ending poverty in all its forms everywhere by 2030 \cite{dehury2015regional}, \cite{alkire2011understandings}, \cite{alkire2013multidimensional}, \cite{alkire2015multidimensional}.

In this paper, we consider the multidimensional synthesis poverty measure introduced by \cite{deepawansa2022innovative}. They selected dimensions based on popular literature and keeping the Goal 1 of the SDGs in consideration. The three main dimensions that they considered are Material Deprivation (MD), Deprivation of Social Dimension (SD) and Deprivation of Human Capital (HC). Material deprivation includes housing facilities, consumer durables, and clothing. Deprivation of social dimension considers social network, dignity, and autonomy. Deprivation of human capital covers education, employment, health, and nutrition. 

We devise a model and methodology to gauge the relative impacts of various dimensions of poverty on an individual's multidimensional poverty status within small areas. Additionally, we introduce a model-driven approach for estimating the poverty status of small areas using a composite measure of multidimensional poverty. The use of small area modeling in the context of poverty assessment is commonly known as {\em Poverty Mapping}. In these granular levels, there is usually limited or no information available from surveys, making it essentially a small area problem. Numerous papers, including those by \cite{fay1979estimates}, \cite{battese1988error}, \cite{ha2014methods}, and \cite{das2025approximateHB}, among others have discussed the challenges of small area estimation and potential solutions. However, despite the existing literature on poverty mapping, there is currently no study that specifically addresses the estimation of relative contributions from different dimensions of poverty, utilizing careful small area modeling tools and techniques. 
This paper aims to bridge that gap and contribute to the poverty mapping literature by addressing these important aspects.

In order to tackle these challenges, we adopt approaches that involve incorporating area-level auxiliary information from reliable external sources. \cite{kim2015small} proposed small area modeling combining several sources of data.  \cite{erciulescu2018chapter} discuss the availability of trustworthy area-level auxiliary data for accurate estimation at granular levels (small areas). 
We employ multivariate joint-modeling of scores calculated from different dimensions to effectively estimate the relative contributions of these dimensions to small area level poverty. This allows us to understand the relative importance of each dimension in measuring poverty. Moreover, we develop a model for the multidimensional composite measure, which enables us to generate poverty estimates, in terms of proportions of impoverished people, for small areas. \cite{larsen2003estimation} demonstrate modeling approaches for estimating proportions for small areas, using empirical Bayes (EB) approach. In contrast, we use hierarchical Bayesian (HB) approach for estimating such proportions. By following this comprehensive approach, we aim at deriving more reliable and insightful poverty assessments at  granular levels.

The rest of the paper is structured as follows: we start by briefly describing the data set that we have used in this paper in Section \ref{sec: data desc}. Section \ref{sec: rel-con} discusses the necessary small area model and method for measuring relative contributions from different dimensions of poverty. Section \ref{sec: mult-pov} explores model/s to estimate multidimensional composite measures and associated uncertainty measures for small areas. In Section \ref{sec: data analysis}, we present results obtained from applying the proposed models and methods to the Sri Lankan data. Moreover, we present key findings from these applications. Finally, we conclude the paper by summarizing our findings and discussing potential avenues for future research and extensions.

\section{Data Description}\label{sec: data desc}
The primary data for the analysis conducted in this research, aiming to identify impoverished individuals residing in the Uva province of Sri Lanka, is based on the ``Synthesis method" proposed by \cite{deepawansa2022innovative}. We prepared the set of aggregated-level covariates for our modeling, from the Sri Lankan 2012 census data.

\subsection{Primary survey data}
The authors \cite{deepawansa2022innovative} designed both the sample and questionnaire for a household survey conducted in the Uva province of Sri Lanka. This survey employed a well-structured, pre-tested questionnaire consisting of close-ended questions tailored to the study's objectives. It was carried out over the period of November 2016 to January 2017.

The survey utilized a stratified two-stage sample design across two districts within the Uva province: Badulla and Moneragala. These districts were further divided into five strata, representing different regions. Badulla has the urban, rural and estate regions; and Moneragala has the rural and estate regions. Within these regions, Primary Sampling Units (PSUs) were selected based on a probability proportional to size (PPS) scheme. The total sample size of 1200 housing units was allocated across the strata, with 120 PSUs chosen to represent various socio-economic groups within the province. Of these, 78 PSUs were from the Badulla district, while 42 were from the Moneragala district.

At the second stage of sampling, Secondary Sampling Units (SSUs) were systematically selected from each PSU. Ten housing units (SSUs) were systematically selected from each selected census block (PSU) for inclusion in the survey. All households within these selected housing units were surveyed, with information collected through face-to-face interviews. The survey collected data on the general characteristics of individuals within households. The study's reference population comprised all individuals aged eighteen and above residing in the Uva Province of Sri Lanka, with the individual who responded during the interview serving as the unit of analysis. The selection of respondents within households was not based on any specific method. Respondents were typically individuals over the age of 18 residing in the household. The survey aimed to enumerate all households within the selected housing units, and face-to-face interviews were conducted to collect information on the general characteristics of individuals within households.

The survey targeted a total of 1193 respondents, out of which 730 were female and 463 were male. The sample aimed to be representative of all socio-economic groups within the Uva province. However, for the analysis conducted in this research, only records with complete information on the selected variables (for creating a multidimensional measure) were considered. As a result, the analysis utilized information from 731 out of the 1193 respondents. Note that, the weighting process reflects adjustments due to unit non-responses, as well as item non-responses. Finally, the weights were calibrated to the small area population sizes obtained from the census data.

For the analysis, a binary response variable indicating poverty status (1 for poor, 0 for non-poor) was derived from the  composite multidimensional measure. They considered multiple indicators of poverty, combining them into three distinct dimensions. Based on the scores and cut-offs for these 3 mutually exclusive and exhaustive dimensions, they created their final composite measure. Using a separate cut-off again, they finally created the binary variable, accounting for the person level poverty status. This variable was employed to estimate poverty at lower administrative divisions using Small Area Estimation (SAE) modeling techniques. In the sections that follow, we will talk about certain challenges we faced, even after careful design and collection of the data by the said authors. 

\subsection{Auxiliary data}
Small area literature talks about the importance of the availability of good administrative records, suitable for the response variable. Having these different kinds of data, possibly from different levels, such as area level aggregates or summaries, or unit-level information, increases the flexibility of the hierarchical models \cite{rao2015small}. The auxiliary variables ($x$) used here in this study, were derived from the 2012 Census of Population and Housing conducted by the Department of Census and Statistics of Sri Lanka. These variables included ``floor", ``ownsMobile", ``waterSafe", ``ownsTV", ``wall", and ``toilet", representing percentages of households with specific characteristics aggregated to Divisional Secretary's Divisions (DSDs) within the Uva province. ``Floor" represented the percentage of households with permanent flooring materials, ``ownsMobile" indicated the percentage with mobile phone ownership, ``waterSafe" represented households with access to safe drinking water, ``ownsTV" indicated households owning a television, ``wall" represented households with permanent wall materials, and ``toilet" indicated households with access to improved sanitation facilities. For more details on how the different indicators were chosen, how they were combined into the different dimensions, and the details about the selection of cut-offs, readers are encouraged to read their work \cite{deepawansa2022innovative}.

\section{Relative Contributions from Different Dimensions of Poverty: Notations, Definitions, and Modeling}\label{sec: rel-con}

In this section, we delve into the intricate dimensions of poverty and explore their relative contributions towards the small areas' poverty landscape. By establishing clear notations, defining key concepts, and formulating appropriate models, we aim to gain deeper insights into the multifaceted nature of poverty and its underlying determinants. Through rigorous analysis and modeling techniques, we seek to unravel the complex interplay between various poverty dimensions and their impacts on individuals and communities.

In our study, we adopt a multidimensional approach to measure poverty, as outlined in \cite{deepawansa2022innovative}. This approach, known as the `synthesis method,' integrates diverse poverty indicators into distinct `dimensions.' By amalgamating these dimensions, the method generates a comprehensive assessment of poverty, represented by a singular score. Through the application of a predefined threshold, the authors dichotomized the composite measure, assigning a value of 1 to individuals classified as poor and 0 to those deemed non-poor. The dimensions identified in their study include `material deprivation,' `social deprivation,' and `human capital.' For detailed insights into these dimensions, their corresponding scores, the designated thresholds, and the methodology employed for their determination, we direct readers to the original work by \cite{deepawansa2022innovative}.

During our exploration of modeling techniques for the composite measure, aimed at making inferences for smaller regions within the Uva province, a pertinent query emerged: Is it feasible to jointly model the various dimensions, represented by their assigned scores? Such an approach holds the potential to address a crucial scientific inquiry regarding the relative contributions of each dimension to the individual-level poverty status observed across the small areas within the Uva province.

\subsection{Notations}

We will use of the following notations. For $i=1,\cdots,m; \;k=1,\cdots,K$, define
\begin{itemize}
    \item $y_{ijk}$: the dimensional score for the $k^{th}$ dimension of poverty for the $j^{th}$ person in the $i^{th}$ small area. 
    \item $\vec\theta_i=(\theta_{i1},\cdots,\theta_{iK})' = $ vector of `true' small area scores, of different dimensions of poverty, for area $i$. Also, $\theta_{i1}, \ldots, \theta_{iK}$ are all greater than 0 and less than 1. This will be our small area parameter of interest, which we aim to estimate through the use of modeling.
    \item ${y}_{ik}$: survey-weighted  estimate of $\theta_{ik}$; denote $\vec{y}_i=({y}_{i1}, \ldots, {y}_{iK})'$. We use the survey design to calculate these estimates from the person level data.
    
    \item $\Sigma_i=((\sigma_{i;kk'}))$:  a known $K\times K$ sampling covariance matrix of $\vec{y}_i$. $\sigma_{i; kk'}$ is the sampling covariance between ${y}_{ik}$ and ${y}_{ik'}$. In practice, $\Sigma_i$ is often estimated using a smoothing technique. 
\end{itemize}
We outline the estimation procedure of $\Sigma_i$ later in Section \ref{sec: mult-pov}.

\subsection{Model}
With the notations set forth, we are equipped to devise a model aligned with our research objective: discerning the relative contributions of various dimensions of poverty on the multidimensional poverty for small areas. Given that the dimensional scores are uniformly 0 for non-poor individuals, our model will focus solely on the impoverished population. In what follows, we present a hierarchical model that establishes a link between the true dimensional scores of small areas and the available area-level covariates. While multivariate hierarchical models have been employed in small area research previously, as seen in works like \cite{datta1991hierarchical} and \cite{benavent2016multivariate}, our approach differs significantly. Unlike these studies, our model incorporates unmatched level 2 linking models, and our parameter of interest varies. Moreover, to the best of our knowledge, this specific type of multivariate unmatched model has not been utilized in the context of estimating relative contributions of various dimensions in multidimensional poverty mapping. Hence, we introduce the following Multivariate Normal Hierarchical Logistic model. The proposed model can be considered as a multivariate extension of the unmatched sampling and linking model, considered by \cite{you2002small} and \cite{sugasawa2018small}. For small areas $i=1,\cdots,m$,
\begin{align}\label{eqn:mn-rs}
    \mbox{Level 1 (sampling model):} \quad \quad \quad \quad \ \  {y}_i|\theta_i & \stackrel{ind}{\sim} MN(\theta_i, \Sigma_i), \nonumber \\
    \mbox{Level 2 (linking model):} \quad \mbox{logit}(\theta_i) | \beta, A & \stackrel{ind}{\sim} MN(X_i\beta, A),
\end{align}
where $\{{y}_i; \ i = 1, \ldots, m\}$ is a $K\times 1$ vector of direct survey-weighted estimates of  $\theta_i=(\theta_{i1},\cdots,\theta_{iK})'$,  a $K\times 1$ vector of true means of $K$ dimensional scores; MN denotes a Multivariate-Normal distribution; $\Sigma_i$ is a $K\times K$ matrix of known sampling covariance matrix of $y_i$; $X_i$ is a $K\times p$ matrix of known auxiliary variables for area $i$;  $\beta$ is a $K\times 1$ vector of unknown regression coefficients;  $A$ is a $K\times K$ matrix of unknown model covariance matrix. We estimate $\beta$ and $A$ from the model, given data. Note that the linking-model facilitates choice of different sets of auxiliary variables possibly from reliable external sources, different from the primary sample survey for different dimensions.
\subsection{Contribution of different dimensions}
For the $i^{th}$ area, we define the contribution of the $k^{th}$ dimension to the overall poverty as $\eta_{ik}$:
$$\eta_{ik} = \frac{\theta_{ik}}{\theta_{i1} + \cdots + \theta_{iK}}.$$ This is our parameter of interest.

\section{Multidimensional Poverty Rate: Modeling, and Estimation} \label{sec: mult-pov}

In the preceding section, we delved into the intricacies of modeling scores derived from various dimensions of poverty, unveiling insights into the relative contributions of these dimensions to an area's overall poverty. Building upon this foundation, we now shift our focus to the modeling of a composite score, which synthesizes the diverse dimensions of poverty into a unified measure. Such modeling endeavors enable the estimation of small area means for person-level poverty, accompanied by robust assessments of estimation accuracy through standard errors. As with our previous exploration, we commence by defining certain quantities of interest that helps with the formulation of the model, and conclude with a discussion on the estimation of population parameters within this framework. Through this approach, we aim to illuminate the methodology underlying the estimation of Multidimensional Poverty Rates (MPRs), fostering a deeper understanding of poverty dynamics at the micro level.


\subsection{Model}
Let $s_i$ be the household-level sample available for the $i^{th}$ small area. For small areas $i = 1, \ldots, m$, define: $\vec{z} = (z_1, \ldots, z_m)'$, where $z_i$ is the estimated proportion of impoverished people in area $i$. $z_i$ is the direct design based estimator of $\pi_i$, which is the true proportion of impoverished people in the $i^{th}$ area.
    \begin{equation}\label{eqn: z_i direct}
    z_i = \frac{\sum\limits_{h, j \in s_i}w_{ihj}z_{ihj}}{\sum\limits_{h, j \in s_i}w_{ihj}}, i = 1, \ldots, m, \, h = 1, \ldots, n_i, \ \text{and} \ j = 1, \ldots, m_h.
\end{equation} where $z_{ihj}$ is the poverty status of the $j^{th}$ person, within the $h^{th}$ household (HH), within the $i^{th}$ small area. $z_{ihj} = 1$, if the said person is poor, and 0 otherwise. $w_{ihj}$ is the corresponding person-level survey weight. Naturally, $n_i = |s_i|$, is the number of HHs in the sample for the small area $i$ and $m_h$ is the HH size of the $h^{th}$ HH.

Now, we can formulate a model that characterizes the true means across the $m$ small areas, linking them to area-level auxiliary information $x_i$, through hierarchical structures. Traditional area-level models, such as the Fay-Herriot model (\cite{fay1979estimates}), and its various extensions, including unmatched level 2 variations, serve as suitable candidates for this purpose. For small areas $i = 1, \cdots, m$, the FH model is given by:

\begin{align}\label{model: FH}
    \mbox{Level 1 (sampling model):} \quad \ \ \  z_i|\pi_i & \stackrel{ind}{\sim} N(\pi_i, D_i) \nonumber \\
    \mbox{Level 2 (linking model):} \quad \pi_i | \gamma, \sigma^2_v & \stackrel{ind}{\sim} N(x_i'\gamma, \sigma^2_v)
\end{align} where $x_i'$ is a vector of known covariates available for the $m$ small areas. $D_i$'s are the known sampling variances. The hyperparameters $\gamma$ and $\sigma^2_v$ are unknown. The problem with this model is that the assumption for normality cannot guarantee that the support stays between 0 and 1. Likewise, the normality assumption does not guarantee the support between 0 and 1 for the posterior distribution of $\pi_i$. The logit extension of the Fay-Herriot model, often referred to as the Normal-Logistic model, incorporates logistic transformations of the parameters to better handle binary and proportion data. The Normal-Logistic (NL) model can be considered by changing the level 2 under the FH model in equation (1). For small areas $i = 1, \cdots, m$

\begin{align}\label{mod: NL}
    \mbox{Level 1 (sampling model):} \quad \quad \quad \quad \ \  z_i|\pi_i & \stackrel{ind}{\sim} N(\pi_i, D_i) \nonumber \\
    \mbox{Level 2 (linking model):} \quad \mbox{logit}(\pi_i) | \gamma, \sigma^2_v & \stackrel{ind}{\sim} N(x_i'\gamma, \sigma^2_v)
\end{align} For standard implementation of the FH (equation \ref{model: FH}) and NL (equation \ref{mod: NL}) models, the sampling variances $D_i$ are estimated using additional design information (within small areas), but errors due to estimation of sampling variances are generally ignored in the subsequent inferences. Drawing inspiration from the methodology outlined by \cite{ha2014methods}, we adopt a similar approach. Moreover, to effectively model the sampling variances ($D_i$) associated with the direct estimates in terms of design effect, we propose the Normal-Logistic Random Sampling Variance (NL$_{rs}$) model. Specifically, for small areas indexed by $i = 1, \ldots, m$, the model takes the following form:
\begin{align}\label{eqn:rs}
    \mbox{Level 1 (sampling model):} \quad \quad \quad \quad \ \  z_i|\pi_i & \stackrel{ind}{\sim} N(\pi_i, D_i = \frac{\pi_i (1 - \pi_i)}{\tilde{n}_i} \mbox{\text{DEFF}}), \nonumber \\
    \mbox{Level 2 (linking model):} \quad \mbox{logit}(\pi_i) | \gamma, \sigma^2_v & \stackrel{ind}{\sim} N(x_i'\gamma, \sigma^2_v),
\end{align}
where $\gamma$ and $\sigma^2_v$ are unknown model parameters, to be estimated from the model, given data.
In model \ref{eqn:rs}, $\text{DEFF}$ represents the true design effect, given by: 
$$\text{DEFF} = \frac{\text{Var}_{{\text{design}}}}{\text{Var}_{\text{SRS}}},$$ where $\text{Var}_{\text{design}}$ is the true sampling variance of the survey-weighted estimate $z_i$, of $\phi_i$, under the given complex sampling design. While $\text{Var}_{\text{SRS}}$ is the true sampling variance of an unweighted estimate under Simple Random Sampling (SRS) of size $n_i$ with replacement. $\tilde{n}_i$ is the ‘adjusted’ sample size of individuals for the area $i$. We will shortly explain this in detail. This adjustment was done on top of calculating the DEFF. Finally, we considered the Normal-Logistic plugin model ($\text{NL}_{plugin}$). For this model, the $\pi_i$ in the random sampling variance model in equation \ref{mod: NL}, is replaced by the Bayes estimate of $\pi_i$.

Let us now explore the quantity used in the first level of model \ref{eqn:rs}: $\tilde{n}_i$. Traditionally, sampling variances of the direct estimators, $D_i$, if modeled, are often expressed as:
\begin{equation}\label{eqn:usual D_i}
    D_i = \frac{\pi_i (1 - \pi_i)}{n_{i; \text{eff}}} = \frac{\pi_i (1 - \pi_i)}{n_i}\text{DEFF} 
\end{equation} where DEFF represents the design effect and $n_i$ is the sample size for the small area $i$ under simple random sampling. We calculated the DEFF from the entire dataset. This captures the complexity of the sampling design. The form given above assumes simple random sampling of unit-level samples, and then corrects it for the complex sampling design by multiplying the variance under SRS by the DEFF. In the context of a complex sampling design with $n$ samples, $n_{\text{eff}} = \frac{n}{\text{DEFF}}$ is often referred to as the effective sample size (ESS). This concept proves useful in evaluating the amount of independent information obtained from a complex survey sample. We have used a different form of sampling variance:
\begin{equation}\label{eqn:our D_i}
    D_i = \frac{\pi_i (1 - \pi_i)}{\tilde{n}_i}\text{DEFF}
\end{equation} as opposed to the usual form given in equation \ref{eqn:usual D_i}. This approach was inspired by the data collection methodology employed by \cite{deepawansa2022innovative} in their study. In their survey, household poverty status aligns with the person-level poverty status for all individuals within the same household. Hence, the sample size for an area $i$ essentially represents the number of households in the survey data available for that area $i$. Since our parameter of interest, $\pi_i$, denotes the proportion of impoverished people for the $i^{th}$ small area, we are specifically interested in person-level poverty. Therefore, in level 1 of equation \ref{eqn:rs}, we adjusted this household-level $n_i$, for the person-level sample size, denoted as $\tilde{n}_i$.

It is worth noting that despite these considerations, the survey weights associated with individuals remain valid due to the complete enumeration conducted within the SSUs. Therefore, the inclusion probabilities, calculated up to the SSUs, ensure that these individuals retain consistent weights. Additionally, the final weights were calibrated to the small area population sizes obtained from the census data, as determined by the authors.

\subsubsection[container]{Calculation of $\tilde{n}$}
We now discuss the calculation of the adjusted sample size, $\tilde{n}$, for any small area. Recall that, $z_i$ given in equation \ref{eqn: z_i direct}, is the direct survey-weighted estimate based on $s_i$, of $\pi_i$. We want to calculate the variance of this estimator $z_i$ under the complex sampling. We drop the small area index $i$ in the following steps, for ease of explanation.

Initially, we compute the variance under the simple random sampling (SRS) framework. Subsequently, we adjust this variance by multiplying it with the design effect (DEFF) to account for the complexity of the sampling design. For instance, suppose $n$ households were sampled using the SRS scheme. Dropping the small area index $i$, let $z_h$ represent the poverty status at the household level, where $z_h = 1$ indicates that the HH is poor and 0 otherwise. This definition aligns with that of $z_{ihj}$ in equation \ref{eqn: z_i direct}. Recall that all individuals residing within a HH $h$ have the same poverty status, which in turn is labeled as the HH-level poverty status. Thus, we have that $z_{hj} = z_h, \ \forall j$ within that household $h$ (dropping the small area index $i$ uniformly). Now, under SRS, with equal weights, we can express equation $\ref{eqn: z_i direct}$ (again dropping the small area index $i$ uniformly) as: 
\begin{equation}\label{eqn: z_i SRS}
    z = \frac{\sum\limits_{h, j \in s} z_{hj}}{\sum\limits_{h, j}1} = \frac{\sum\limits_{h = 1}^{m_h} m_h z_h}{\sum\limits_{h = 1}^{m_h} m_h}, \ \text{since} \ z_{hj} = z_h, \ \forall j.
\end{equation} Under SRS, we can assume that HHs were sampled from a superpopulation of households with $z_h \stackrel{iid}\sim B(\pi)$, where $B$ is the Bernoulli distribution. Then, under SRS we can obtain the variance of $z$ as:
\begin{equation}\label{eqn: var(z)}
    \text{Var}(z) = v(z) \stackrel{iid}= \frac{\sum\limits_{h = 1}^{m_h}m^2_h v(z_h)}{(\sum\limits_{h = 1}^{m_h} m_h)^2} =  \frac{\sum\limits_{h = 1}^{m_h}m^2_h \pi (1 - \pi)}{(\sum\limits_{h = 1}^{m_h} m_h)^2} = \frac{\pi (1 - \pi)}{\tilde{n}}, \ \text{say.}
\end{equation} From the above equation \ref{eqn: var(z)}, we can obtain the adjusted $\tilde{n}$ as: $$\tilde{n} = \frac{(\sum_{h=1}^n m_h)^2}{\sum_{h=1}^n m_h^2}.$$ If $m_h = m, \ \forall \ h$, then $v(z) = \frac{\pi (1 - \pi)}{n}$. We will have $\frac{\tilde{n}}{n} \approx 1$, if the HH sizes do not vary much. Using Cauchy-Schwarz inequality, we can show that $\tilde{n} \leq n$. We can interpret $\tilde{n}$ as: 
\begin{equation*}
    \tilde{n} = \frac{n}{\Delta},
\end{equation*} where $\Delta$ is the adjustment factor for estimating the $v(z)$ under person-level poverty, instead of HH-level poverty. $\Delta$ can be interpreted as the ratio of the variance of the estimator under HH-level poverty, to the variance of the estimator under the person-level poverty. Similarity in defining $n_{\text{eff}}$, with this adjustment factor is noteworthy. We can follow the same steps for calculating $\tilde{n}_i$ for a particular small area $i$.

We will now discuss the estimation of $\Sigma_i$ in model \ref{eqn:mn-rs}. We delayed this discussion because the estimation steps rely on the calculations of $\tilde{n}_i$, which we had just described above. Recall that, $\Sigma_i=((\sigma_{i;kk'}))$. We propose the following smoothed estimate of $\sigma_{i; kk}$: $$\hat{\sigma}_{i; kk} = \widehat{var}({y}_{ik}) = \frac{s_{kk}}{n_{i,k; \text{eff}}},$$ 
where $n_{i,k; \text{eff}} = \frac{\tilde{n}_i}{\text{DEFF}_k}$ is the effective sample size of the $k^{th}$ dimension in the $i^{th}$ area; $s_{kk}$ is the pooled variance of $y_{ijk}$ and is defined as: $$s_{kk} = \frac{\sum_{i, j}(y_{ijk} - \bar{y}_{..k})^2}{P - 1}, \ \bar{y}_{..k} = \frac{\sum_{ij}y_{ijk}}{\sum_{i, j}1},  \ P = \text{\# poor individuals in the sample}.$$ Note that, the dimensional scores for the non-poor individuals are 0. The calculation of the $\tilde{n}_i$ will follow the exact same method steps as we discussed above. $\text{DEFF}_k$ is the design effect corresponding to the $k^{th}$ dimension of poverty. To smooth the covariance terms, $\sigma_{i; kk'}$, consider the following identity: $$var({y}_{ik} + {y}_{ik'}) = var({y}_{ik}) + var({y}_{ik'}) + 2 cov({y}_{ik}, {y}_{ik'}).$$
Using the above, we can get smoothed estimates of the $\sigma_{i; kk'}$ as:

$$\hat{\sigma}_{i; kk'} = \widehat{cov}({y}_{ik}, {y}_{ik'}) = \frac{\widehat{var}({y}_{ik} + {y}_{ik'}) - \widehat{var}({y}_{ik}) - \widehat{var}({y}_{ik'})}{2}.$$ 
To smooth $var({y}_{ik} + {y}_{ik'})$ we follow the method used to smooth $var({y}_{ik})$.  But, we consider the new variable: $(y_{ijk} + y_{ijk'})$ instead of just $y_{ijk}$.

\subsection{Hierarchical Bayesian Estimation}

In line with typical Hierarchical Bayesian (HB) methodology, we employ Markov Chain Monte Carlo (MCMC) methods to estimate the model parameters. Let $\theta_i$ represent the population parameter of interest for area $i$. Subsequently, we utilize the set of $R$ MCMC replicates of $\theta_i$, denoted as $\{\theta_i^{(r)}, r = 1, \ldots, R\}$, for making inferences regarding $\theta_i$, the true population parameter. For the Multivariate Normal Hierarchical Logistic model (see Section \ref{sec: rel-con}), we parameterize matrix $A$ as follows:
\[A = \sigma^2  \begin{pmatrix}
1 & \rho_{12} & \dots & \rho_{1k} \\
 & 1 & \dots & \rho_{2k} \\
 &  &  & 1 
\end{pmatrix}\]
Here, $\sigma^2$ denotes the common variance component, and $\rho_{i; kk'} = \text{corr}(\text{logit}(\theta_{ik}), \text{logit}(\theta_{ik'}))$ for all small areas $i$ and dimensions $(k, k')$. We adopt weakly informative proper priors for the components of matrix $A$, specifically: $\sigma \sim \text{Cauchy}^{+} (0, 5)$, a Half-Cauchy prior \cite{gelman2008weakly}, and the correlation coefficients $\rho_{ik} \stackrel{iid}\sim \text{Uniform} (0, 1)$. Although correlation coefficients could theoretically be negative, \cite{deepawansa2022innovative} suggest they are all positive. Thus, we use a positive uniform prior ranging from 0 to 1 for these correlation entries in matrix $A$. Finally, for each MCMC replicate $r = 1, \cdots, R$, we compute:
\[\eta^{(r)}_{ik} = \frac{\theta^{(r)}_{ik}}{\theta^{(r)}_{i1} + \cdots + \theta^{(r)}_{iK}}\]
This set of $R$ replicates facilitates posterior inference about the contributions.

\subsection{MCMC with Stan} \label{sec: StanMCMC}
Bayesian computations in this study are conducted using $\mathbf{Stan}$ (\cite{StanManual}), a probabilistic programming language employing Hamiltonian Monte Carlo (HMC). HMC efficiently spans the posterior distribution by utilizing the derivative of the posterior density function. Stan integrates an approximate Hamiltonian dynamics simulation based on numerical integration, coupled with a Metropolis-Accept step. Readers are referred to \cite{StanManual} and \cite{gelman2013bayesian} for a comprehensive understanding of HMC and how Stan implements it.

Stan facilitates model comparison via the \texttt{loo} package \cite{vehtari2017practical}, computing approximate Leave-One-Out Cross-Validation (LOO-CV) using Pareto smoothed importance sampling (PSIS). This package also estimates differences in expected predictive accuracy between models through the difference in their $\hat{\text{elpd}}_{\text{loo}}$, the Bayes Leave-One-Out estimate of the expected log pointwise predictive density (elpd). For a large enough sample size, this difference approximates a standard normal distribution \cite{gelman2013bayesian}, facilitating model comparisons.

All models in this study were manually coded in $\mathbf{Stan}$ using its specific programming language due to the absence of suitable packages directly implementing such models. Leveraging $\mathbf{Stan}$'s programming flexibility, we successfully implemented all necessary models. Posterior MCMC draws were saved and imported into $\mathbf{R}$ for posterior inferences and visualizations. Special attention was given to coding to ensure the positive-definiteness of the correlation matrix in equation \ref{eqn:mn-rs}. Posterior distributions were structured during coding to enable the use of the \texttt{loo} package for computing $\hat{\text{elpd}}_{\text{loo}}$ for model comparison purposes. This was the sole instance where external package code was utilized.

\section{Data Analysis}\label{sec: data analysis} 

We carry out all the analyses using the probabilistic programming language $\textbf{Stan}$, along with one of Stan's helper packages for $\textbf{R}$ (\texttt{loo}).

\subsection{Joint modeling of the dimensional scores for estimating relative contributions of three dimensions towards the poverty status}

In this subsection, we report our findings from joint modeling of the three dimensions that were used to create the final composite measure of poverty status. Our goal here is to attribute different dimensions towards the final poverty status of the small areas.

For our dataset of consideration, we did not have access to dimension-specific auxiliary variables. So, for our case, $X_i\beta =  \vec{x}_i'\beta 1_K$, where $1_K$ is the unit vector of dimension $K$. As already discussed in Section \ref{sec: rel-con}, the joint modeling of the 3 dimensions requires data only for impoverished people as determined by the composite score and the `Synthesis Method'. We note that for the small area `Meegahakivula', there is only one sample with a `not poor' poverty status. For this particular area, we use our model and the posterior estimates of the model parameters to create the contribution scores for this particular small area. This, again, demonstrates one merit of the small area modeling techniques. With the help of auxiliary variables from Census, and along with the MCMC posterior samples of the model parameters, we could provide the estimate (with the desired uncertainty statistics) even for this small area with no sample from the primary survey data. We restricted all prior choices to the proper Weakly Informative Priors (WIP) \cite{gelman2008weakly}. The hyperparameters at Level 2 of the Multivariate Normal Hierarchical Logistic model \ref{eqn:mn-rs} are given the following priors: $\beta_p \stackrel{ind}\sim N(0, 5^2), \ p = 1, \cdots 7$, $\sigma \sim \text{Cauchy}^{+}(0, 5)$, and $\rho \sim \text{Uniform}(0, 1)$. We run each model in 4 parallel chains, with $10,000$ iterations in each chain.  We discard the first $5,000$ as burn-in samples from each chain. Thus, all the posterior analyses are based on $20,000$ post-warm-up MCMC draws. 

Table \ref{contributions} displays estimates of the contributions of all the 3 dimensions for all the 26 small areas. From this table, we observe that for all the 26 areas, `Social Deprivation' (sd) dimension contributes the most to an area's person-level poverty status while each of the other two dimensions—`Material Deprivation' (md), and `Human Capital' (hc)—has about equal contribution to the same. In regions like the Uva province in Sri Lanka, there is limited diversification in economic and infrastructure development among its Divisional Secretariate (DS) divisions. Predominantly agrarian, Uva is considered one of the remotest and economically lagging areas compared to more developed provinces like Western and Central Sri Lanka. The 2012 census of population and housing evidenced minimal diversity in housing structures and household possessions such as televisions and mobile phones among DS divisions (DCS, 2012, \url{http://www.statistics.gov.lk/PopHouSat/CPH2011/index.php?fileName=Activities/TentativelistofPublications}). This lack of diversity is mirrored in the findings of our study, which show minimal disparities in the contributions of different poverty dimensions among the DS divisions. From the dataset we considered, we did not observe significant differences in the contributions of the various dimensions of poverty to the overall poverty statuses of the small areas. This finding aligns with the results of the Sri Lankan census report within the Uva province. Nonetheless, we are confident that our methodology is capable of accurately reflecting the reality of contributions from different dimensions in other datasets, possibly with different sets of dimensions creating the multidimensional poverty index.

\begin{table}[] \centering 
  \caption{Bayes estimates of relative contributions of three dimensions—`Material Deprivation' (md), `Social Deprivation' (sd) and `Human Capital' (hc)—for all the 26 small areas from the Uva province of Sri Lanka. The table also gives the standard errors for the estimates (denoted by md\_se, sd\_se and hc\_se) along with their corresponding 95\% Bayesian Credible Intervals (\_2.5\%, \_97.5\%) for each dimension for 26 small areas.} 
  \label{contributions} 

\begin{adjustbox}{width=\textwidth} 
\begin{tabular}{@{\extracolsep{5pt}} cccccccccccccc} 
\\[-1.8ex]\hline 
\hline \\[-1.8ex] 
 & small areas & md & md\_se & md\_2.5\% & md\_97.5\% & sd & sd\_se & sd\_2.5\% & sd\_97.5\% & hc & hc\_se & hc\_2.5\% & hc\_97.5\% \\ 
\hline \\[-1.8ex] 
1 & Badalkumbura & $0.242$ & $0.010$ & $0.224$ & $0.263$ & $0.470$ & $0.011$ & $0.447$ & $0.491$ & $0.288$ & $0.010$ & $0.269$ & $0.308$ \\ 
2 & Badulla & $0.234$ & $0.010$ & $0.214$ & $0.255$ & $0.476$ & $0.018$ & $0.445$ & $0.516$ & $0.290$ & $0.018$ & $0.245$ & $0.320$ \\ 
3 & Bandarawela & $0.234$ & $0.010$ & $0.217$ & $0.256$ & $0.486$ & $0.012$ & $0.462$ & $0.509$ & $0.280$ & $0.011$ & $0.258$ & $0.300$ \\ 
4 & Bibile & $0.240$ & $0.010$ & $0.220$ & $0.262$ & $0.457$ & $0.013$ & $0.429$ & $0.480$ & $0.303$ & $0.012$ & $0.282$ & $0.326$ \\ 
5 & Buttala & $0.230$ & $0.010$ & $0.210$ & $0.248$ & $0.492$ & $0.012$ & $0.471$ & $0.517$ & $0.278$ & $0.011$ & $0.256$ & $0.297$ \\ 
6 & Ella & $0.250$ & $0.011$ & $0.230$ & $0.272$ & $0.487$ & $0.013$ & $0.463$ & $0.512$ & $0.263$ & $0.011$ & $0.241$ & $0.284$ \\ 
7 & Haldummulla & $0.248$ & $0.010$ & $0.228$ & $0.270$ & $0.490$ & $0.014$ & $0.463$ & $0.517$ & $0.262$ & $0.015$ & $0.234$ & $0.291$ \\ 
8 & Hali-Ela & $0.253$ & $0.011$ & $0.234$ & $0.276$ & $0.482$ & $0.011$ & $0.459$ & $0.505$ & $0.265$ & $0.011$ & $0.242$ & $0.286$ \\ 
9 & Haputale & $0.233$ & $0.009$ & $0.213$ & $0.250$ & $0.499$ & $0.012$ & $0.477$ & $0.525$ & $0.268$ & $0.010$ & $0.247$ & $0.288$ \\ 
10 & Kandaketiya & $0.249$ & $0.010$ & $0.230$ & $0.270$ & $0.449$ & $0.014$ & $0.421$ & $0.475$ & $0.302$ & $0.012$ & $0.280$ & $0.326$ \\ 
11 & Katharagama & $0.227$ & $0.011$ & $0.205$ & $0.247$ & $0.477$ & $0.012$ & $0.454$ & $0.503$ & $0.296$ & $0.011$ & $0.274$ & $0.319$ \\ 
12 & Lunugala & $0.255$ & $0.011$ & $0.232$ & $0.276$ & $0.482$ & $0.017$ & $0.447$ & $0.515$ & $0.263$ & $0.017$ & $0.228$ & $0.293$ \\ 
13 & Madulla & $0.253$ & $0.010$ & $0.235$ & $0.275$ & $0.452$ & $0.014$ & $0.423$ & $0.478$ & $0.295$ & $0.011$ & $0.273$ & $0.319$ \\ 
14 & Mahiyanganaya & $0.222$ & $0.010$ & $0.201$ & $0.241$ & $0.491$ & $0.011$ & $0.468$ & $0.511$ & $0.288$ & $0.011$ & $0.268$ & $0.308$ \\ 
15 & Medagama & $0.242$ & $0.010$ & $0.222$ & $0.261$ & $0.469$ & $0.013$ & $0.444$ & $0.495$ & $0.289$ & $0.012$ & $0.266$ & $0.313$ \\ 
16 & Meegahakivula & $0.262$ & $0.009$ & $0.244$ & $0.281$ & $0.455$ & $0.014$ & $0.427$ & $0.481$ & $0.283$ & $0.012$ & $0.259$ & $0.307$ \\ 
17 & Moneragala & $0.246$ & $0.009$ & $0.230$ & $0.266$ & $0.462$ & $0.011$ & $0.440$ & $0.483$ & $0.292$ & $0.009$ & $0.273$ & $0.310$ \\ 
18 & Passara & $0.250$ & $0.010$ & $0.230$ & $0.272$ & $0.480$ & $0.015$ & $0.452$ & $0.507$ & $0.270$ & $0.016$ & $0.241$ & $0.300$ \\ 
19 & Rideemaliyadda & $0.233$ & $0.011$ & $0.210$ & $0.254$ & $0.500$ & $0.016$ & $0.473$ & $0.532$ & $0.267$ & $0.014$ & $0.239$ & $0.294$ \\ 
20 & Sevanagala & $0.223$ & $0.011$ & $0.201$ & $0.242$ & $0.485$ & $0.012$ & $0.463$ & $0.509$ & $0.293$ & $0.010$ & $0.273$ & $0.314$ \\ 
21 & Siyambalanduwa & $0.262$ & $0.013$ & $0.239$ & $0.288$ & $0.456$ & $0.016$ & $0.423$ & $0.489$ & $0.282$ & $0.015$ & $0.252$ & $0.310$ \\ 
22 & Soranathota & $0.260$ & $0.013$ & $0.238$ & $0.286$ & $0.459$ & $0.014$ & $0.432$ & $0.485$ & $0.281$ & $0.011$ & $0.259$ & $0.304$ \\ 
23 & Thanamalvila & $0.239$ & $0.010$ & $0.218$ & $0.260$ & $0.497$ & $0.016$ & $0.467$ & $0.529$ & $0.264$ & $0.015$ & $0.235$ & $0.291$ \\ 
24 & Uva Paranagama & $0.268$ & $0.014$ & $0.242$ & $0.296$ & $0.443$ & $0.013$ & $0.418$ & $0.468$ & $0.289$ & $0.012$ & $0.267$ & $0.311$ \\ 
25 & Welimada & $0.241$ & $0.009$ & $0.223$ & $0.259$ & $0.480$ & $0.011$ & $0.461$ & $0.503$ & $0.279$ & $0.009$ & $0.260$ & $0.296$ \\ 
26 & Wellawaya & $0.231$ & $0.010$ & $0.211$ & $0.249$ & $0.472$ & $0.010$ & $0.451$ & $0.492$ & $0.298$ & $0.009$ & $0.281$ & $0.318$ \\ 
\hline \\[-1.8ex] 
\end{tabular} 
\end{adjustbox}

\end{table} 

\subsection{Modeling of the multidimensional poverty status to create the small area estimates of poverty for the Uva province}

The main goal of this subsection is to provide the proportions of poor people for each of the 26 small areas. We discuss how the different models performed when they are used to create small area estimates using the real-life data collected and compiled (as discussed in the previous section \ref{sec: data desc}) by \cite{deepawansa2022innovative}. Again, we restrict all prior choices to the proper Weakly Informative Priors (WIP). That is, hyperparameters at Level 2 of the described hierarchical model in \ref{eqn:rs} are given the following priors: $\gamma_p \stackrel{ind}\sim N(0, 5^2), \ p = 1, \cdots 7$, and $\sigma_v \sim \text{Cauchy}^{+}(0, 5)$. We run each model in 4 parallel chains, with $10,000$ iterations in each chain.  We discard the first $5,000$ as burn-in samples from each chain. Thus, all the posterior analyses are based on $20,000$ post-warm-up MCMC draws. 

After fitting all the 3 models, we use a Model Selection criterion to select one `best' working model. The following Table \ref{loo-table} is used to select our working model.


\begin{table}[] \centering 
  \caption{\small{LOO-CV comparison table for four competing models -- $\text{NL}_{rs}$ is the Normal-Logistic Random Sampling model given in Section \ref{sec: mult-pov}; FH is the model proposed by \cite{fay1979estimates} (also see  \cite{ha2014methods}, \cite{liu2007hierarchical}); NL is the univariate normal-logistic model, a special case of model proposed in Section \ref{sec: rel-con} (also see \cite{liu2007hierarchical} and \cite{ha2014methods}); $\text{NL}_{plugin}$ is the random sampling variance model in Section \ref{sec: mult-pov} where $\pi_i$ in $D_i$ is replaced by the Bayes estimate of $\pi_i$. LOO-IC selects $\text{NL}_{rs}$ as the best of the 4 models.} }
  \label{loo-table} 
\begin{adjustbox}{width=0.29\textwidth}
    \begin{tabular}{@{\extracolsep{5pt}} ccc} 
        \\[-1.8ex]\hline 
        \hline \\[-1.8ex] 
         & elpd\_diff & se\_diff \\ 
        \hline \\[-1.8ex] 
        $\text{NL}_{rs}$ & $0$ & $0$ \\ 
        FH & $-0.059$ & $0.226$ \\ 
        NL & $-0.116$ & $0.232$ \\ 
        $\text{NL}_{plugin}$ & $-0.835$ & $0.284$ \\ 
        \hline \\[-1.8ex] 
    \end{tabular} 
\end{adjustbox}

\end{table} 

Refer to subsection \ref{sec: StanMCMC} for the details of the measures used in Table \ref{loo-table}. From this table, we see that LOO information criterion selects  NL$_{rs}$ as the best working model, and it is better than the other 3 models as seen by the \texttt{elpd\_diff}. Now, even though the $\text{NL}_{rs}$ model is not significantly better for the given data, we still prefer this model over the other three, because of the various advantages of having a random-sampling variance component in the model. We discuss such advantages in detail in the following paragraphs.

We had 6 area level covariates from the Census data, as described in section \ref{sec: data analysis}. We start by including all of them as the covariates, but the posterior analysis suggests that not all of them are significant. In Table \ref{summary-full}, we report the Analysis of Variance (ANOVA) table from the full-model fit. The `mean' column gives the posterior means, `se\_mean' column gives the Monte-Carlo standard errors, the `sd' column gives the posterior standard deviations (standard errors of the Bayes estimates), the next 2 columns give the 2 percentiles ($15^{th}$ and $85^{th}$ percentiles respectively), the column n\_eff gives the Effective Sample Sizes (ESS, $n_{\text{eff}}$) -- the number of independent samples that is used to replace the total $n$ dependent MCMC draws having the same estimation power as the $n$ autocorrelated samples. Finally, the column `Rhat' gives a measure of chain-equilibrium ($\hat{R}$) -- a value near 1 (but $< 1.05$) indicates that the chains have mixed well, and the posterior samples can be used with confidence for the posterior analyses (see \cite{gelman2013bayesian} for details). This ANOVA table suggests that only the `floor' and `wall' are significant at 70\%. 

\begin{table}[] \centering 
  \caption{\small{Posterior summary statistics for the Normal-Logistic Random Sampling model parameters of the full model—with all 6 area-level covariates and 1 intercept} }
  \label{summary-full} 
  \begin{adjustbox}{width=0.69\textwidth}
\begin{tabular}{@{\extracolsep{5pt}} cccccccc} 
\\[-1.8ex]\hline 
\hline \\[-1.8ex] 
 & mean & se\_mean & sd & 15\% & 85\% & n\_eff & Rhat \\ 
\hline \\[-1.8ex] 
Intercept & $0.237$ & $0.004$ & $0.167$ & $0.073$ & $0.402$ & $1,649.193$ & $1.007$ \\ 
floor & $$-$0.553$ & $0.018$ & $0.364$ & $$-$0.919$ & $$-$0.203$ & $407.363$ & $1.014$ \\ 
ownsMobile & $$-$0.057$ & $0.012$ & $0.458$ & $$-$0.504$ & $0.395$ & $1,393.808$ & $1.006$ \\ 
waterSafe & $0.023$ & $0.011$ & $0.350$ & $$-$0.341$ & $0.367$ & $1,092.918$ & $1.007$ \\ 
ownsTV & $0.324$ & $0.015$ & $0.455$ & $$-$0.127$ & $0.774$ & $915.239$ & $1.006$ \\ 
wall & $0.304$ & $0.005$ & $0.213$ & $0.093$ & $0.508$ & $1,960.126$ & $1.004$ \\ 
toilet & $0.227$ & $0.009$ & $0.266$ & $$-$0.042$ & $0.496$ & $969.290$ & $1.003$ \\ 
$\sigma^2_v$ & $0.210$ & $0.016$ & $0.290$ & $0.011$ & $0.429$ & $324.210$ & $1.012$ \\ 
\hline \\[-1.8ex] 
\end{tabular} 
\end{adjustbox}

\end{table}

\begin{table}
    \caption{A list of competing models} 
    \label{tab: model_list}
    \begin{center}
    \begin{adjustbox}{width=0.69\textwidth}
    \begin{tabular}{@{\extracolsep{5pt}}  rrrrrrrr}
    \hline 
    \hline 
    Model & Intercept & floor & ownsMobile &  waterSafe & ownsTV & wall & toilet  \\
    \hline
    Full Model & \checkmark & \checkmark& \checkmark  & \checkmark & \checkmark  & \checkmark & \checkmark  \\
    Subset Model & \checkmark   & \checkmark    & &           &   &   \checkmark         &    \\

    \hline
    \end{tabular}%
    \end{adjustbox}
    \end{center}
\end{table}

\begin{table}[] \centering 
  \caption{\small{Posterior summary statistics for the Normal-Logistic Random Sampling model parameters of the subset model—the best Model chosen by LOO criterion with the significant (at 70\%) model }parameters} 
  \label{summary-subset} 
  \begin{adjustbox}{width=0.69\textwidth}
\begin{tabular}{@{\extracolsep{5pt}} cccccccc} 
\\[-1.8ex]\hline 
\hline \\[-1.8ex] 
 & mean & se\_mean & sd & 15\% & 85\% & n\_eff & Rhat \\ 
\hline \\[-1.8ex] 
Intercept & $0.253$ & $0.003$ & $0.158$ & $0.097$ & $0.413$ & $2,894.011$ & $1.001$ \\ 
floor & $$-$0.178$ & $0.003$ & $0.164$ & $$-$0.338$ & $$-$0.013$ & $3,323.882$ & $1.001$ \\ 
wall & $0.128$ & $0.003$ & $0.140$ & $$-$0.012$ & $0.266$ & $2,222.158$ & $1.001$ \\ 
$\sigma^2_v$ & $0.152$ & $0.007$ & $0.202$ & $0.009$ & $0.311$ & $739.094$ & $1.001$ \\ 
\hline \\[-1.8ex] 
\end{tabular} 
  \end{adjustbox}

\end{table} 
Table \ref{tab: model_list} gives a checklist summary of the `Full' model and the `Subset' model. Table \ref{summary-subset} provides the ANOVA table of the subset model with the 2 significant covariates from Table \ref{summary-full}. So we use the only 2 covariates that were significant at 70\% -- `floor' and `wall'. Since LOO selects the $\text{NL}_{rs}$ model, we report another LOO table comparing the $\text{NL}_{rs}$ full model with all 6 covariates with the subset model with the 2 aforementioned covariates. Clearly, from table \ref{loo-table2} we can see that the subset model with 2 covariates is significantly better than the full model with all 6 covariates. 

\begin{table}[!htbp] \centering 
  \caption{\small{LOO-CV comparison table for the `best' model -- $\text{NL}_{rs}$ subset model using only the 2 significant (at 70\%) covariates—with its counterpart using all the 6 small-area level covariates. Both the models include an intercept.} }
  \label{loo-table2} 
  \begin{adjustbox}{width=0.31\textwidth}
      \begin{tabular}{@{\extracolsep{5pt}} ccc} 
        \\[-1.8ex]\hline 
        \hline \\[-1.8ex] 
         & elpd\_diff & se\_diff \\ 
        \hline \\[-1.8ex] 
        $\text{NL}_{rs}$ (subset) & $0$ & $0$ \\ 
        $\text{NL}_{rs}$ (full) & $-2.850$ & $1.909$ \\ 
        
        \hline \\[-1.8ex] 
        \end{tabular} 
  \end{adjustbox}
\end{table}

\begin{figure}[htbp]
    \centering
    \includegraphics[width=0.57\textwidth]{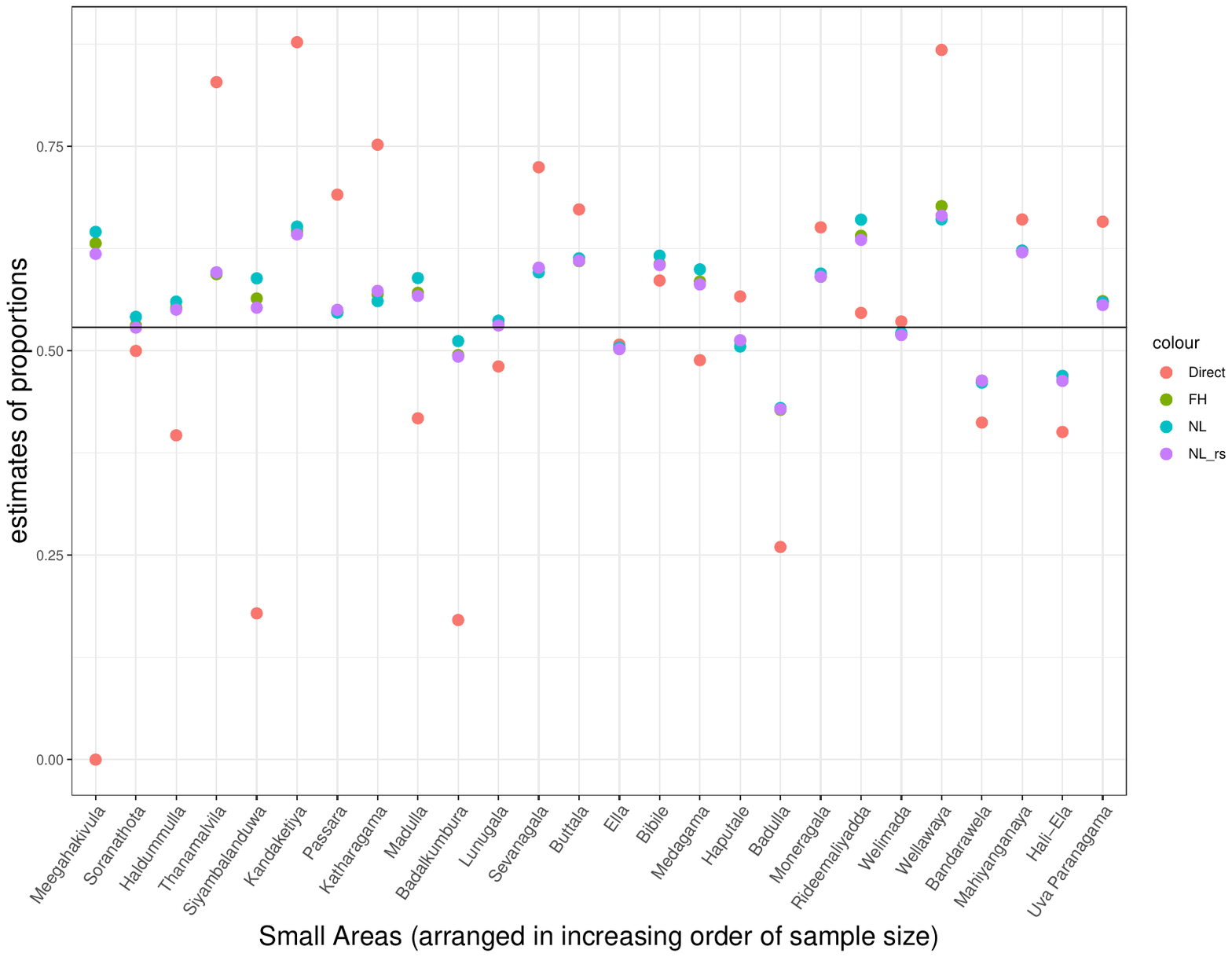}
    \caption{\small{Comparison of direct survey-based estimates (Direct) and 3 Hierarchical model based estimates -- Hierarchical Fay-Herriot (FH), Normal-Logistic (NL) and Normal-Logistic-Random-Sampling (NL$_{rs}$) models. The horizontal line refers to the overall estimate for all the areas. In x-axis, the small areas are arranged increasing order of sample size.}}
    \label{fig:ests-comp}
\end{figure} 
Figure \ref{fig:ests-comp} presents estimates from the 3 hierarchical Bayesian (HB) models (see Table \ref{loo-table}) and the direct survey-weighted estimates for the 26 small areas, organized by increasing sample sizes for each area. The figure also includes the overall estimate across all areas. It illustrates the limitations of direct survey-weighted estimates and the advantages of modeling for obtaining stable estimates. Notably, the model-based estimates alleviate issues seen in certain areas, such as `Meegahakivula', where the direct survey-weighted estimate of 0 may seem implausible due to the limited sample size and the poverty status of the sole individual sampled. Overall, the HB model-based estimates offer reasonable and consistent results, even in areas with limited sample sizes.
\begin{figure}[]
    \centering
    \includegraphics[width=0.53\textwidth]{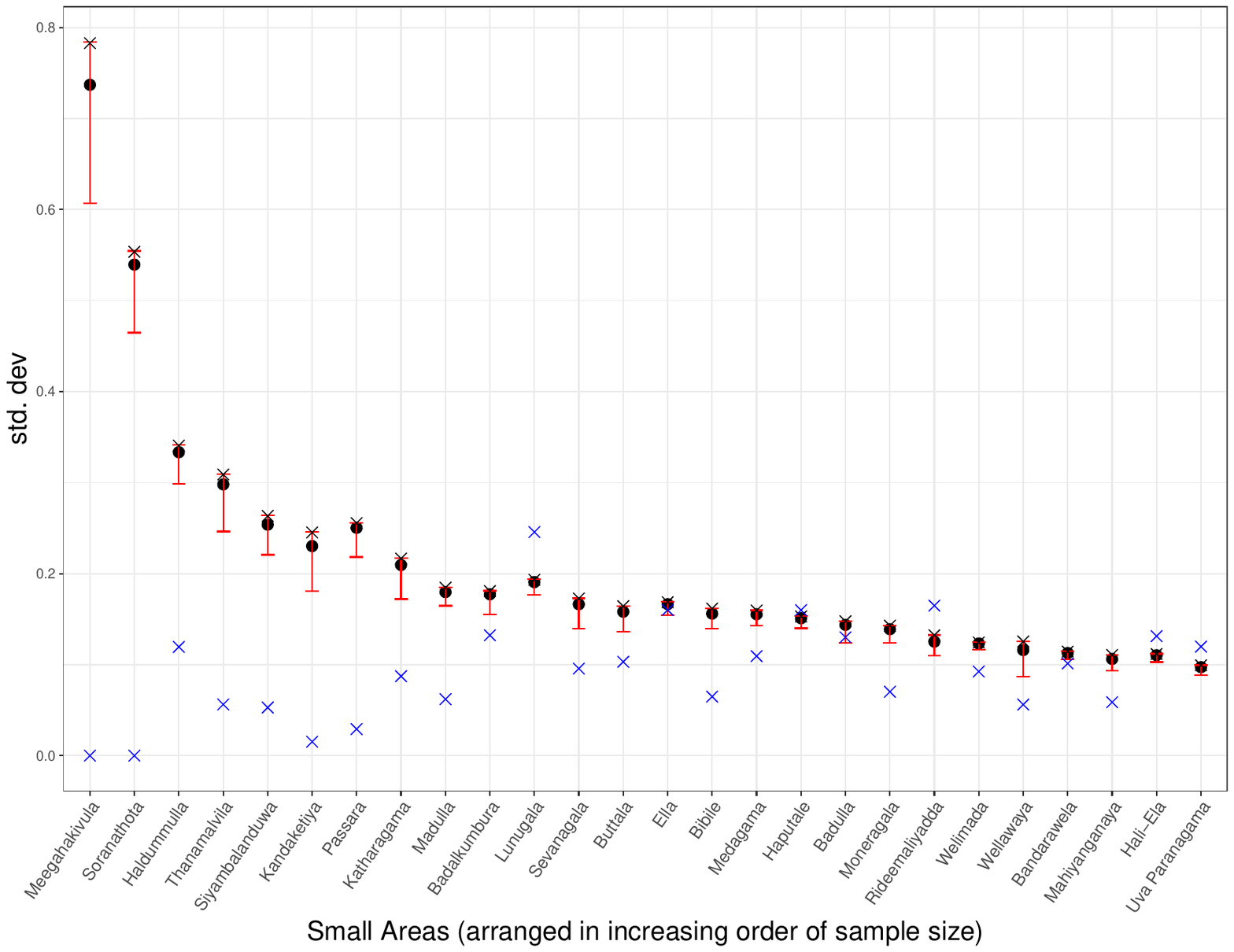}
    \caption{Interval plot of the sampling standard deviations—the red band gives the 95\% posterior Credible Interval of the sampling standard deviations obtained from the NL$_{rs}$  model. The black dot gives the posterior mean of the sampling standard deviation, and the black cross gives the smoothed sampling standard deviations used in the Fay-Herriot model and the Normal-Logistic model. The blue cross gives the design-based standard deviations. In x-axis, the small areas are arranged increasing order of sample size.} 
    \label{fig:int.plot_CI}
\end{figure}

Figure \ref{fig:int.plot_CI} demonstrates  superiority of the NL$_{rs}$ model compared to the other 2 HB models. NL$_{rs}$ provides an idea of variability of the sampling standard deviations when the sample survey data is expected to vary/change. We can observe less variability of the sampling standard deviations for all the small areas over the smoothed design-based standard deviations as were used in the other 2 HB models. Moreover, we can get an estimate of the variability and produce such plots with 95\% posterior credible intervals of the standard deviations. The plot also shows the advantage in using a model-based estimates instead of survey-weighted estimates for small areas. The direct estimates of the standard errors are very unstable, especially for areas with smaller sample sizes.

\begin{figure}[]
    \centering
    \includegraphics[width=0.53\textwidth]{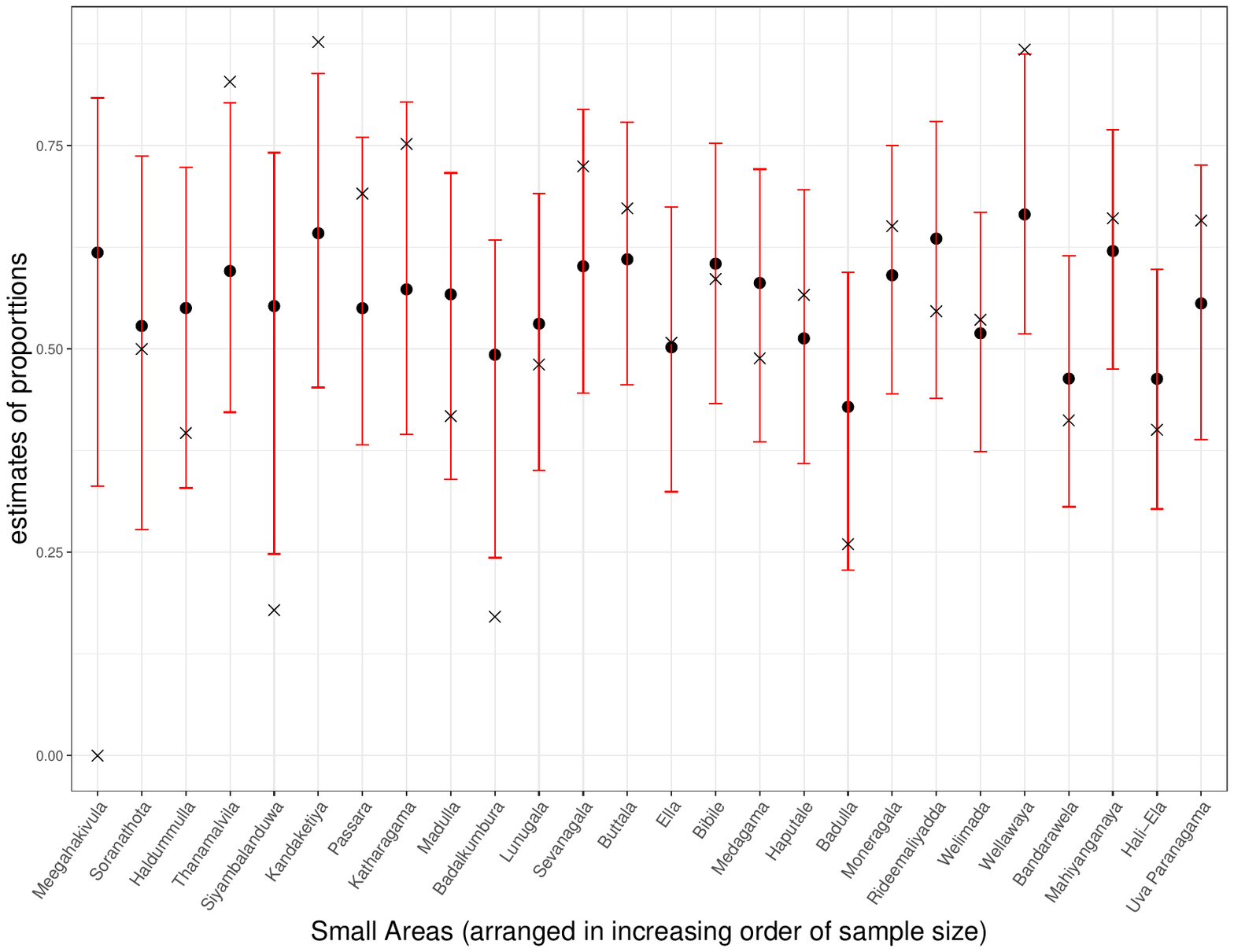}
    \caption{Interval plot of the estimates for 26 small areas -- the red band gives the 95\% posterior Credible Interval of the proportions obtained from the NL$_{rs}$ model. The black dot gives the posterior mean of the proportions, and the black cross gives the survey-based estimates of the proportions. In X-axis, the small areas are arranged increasing order of sample size.} 
    \label{fig:model2_comp_CI}
\end{figure}

Figure \ref{fig:model2_comp_CI} compares small area means, proportions in this case, from the NL$_{rs}$ model (black dots) along with its 95\% Bayesian credible interval and the estimates from the direct survey-weighted method. NL$_{rs}$ not only gives us more believable and reliable estimates for areas with small (or no) samples, it also gives us Bayesian credible intervals of the estimates.

\begin{figure}[]
    \centering
    \includegraphics[width=0.53\textwidth]{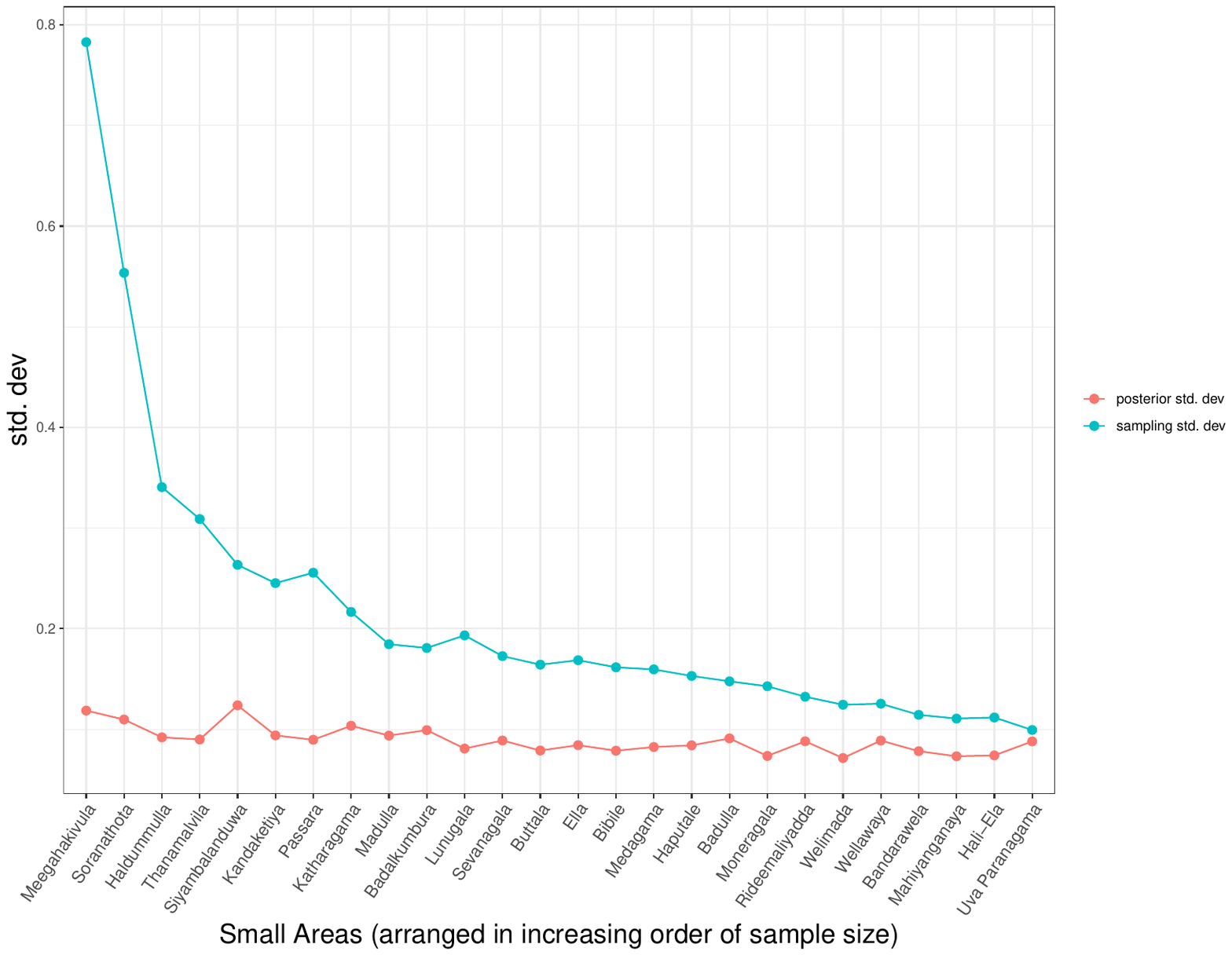}
    \caption{The red line gives the posterior standard deviations of the small area proportions obtained from the `best' fitting NL$_{rs}$ model. On the other hand, the turquoise line gives the smoothed sampling standard deviations ($\sqrt{D_i}$) that were used at the level 1 (sampling level) of the FH and the NL models. In x-axis, the small areas are arranged increasing order of sample size.} 
    \label{fig:psi_samp}
\end{figure}

\begin{figure}[]
    \centering
    \includegraphics[width=0.53\textwidth]{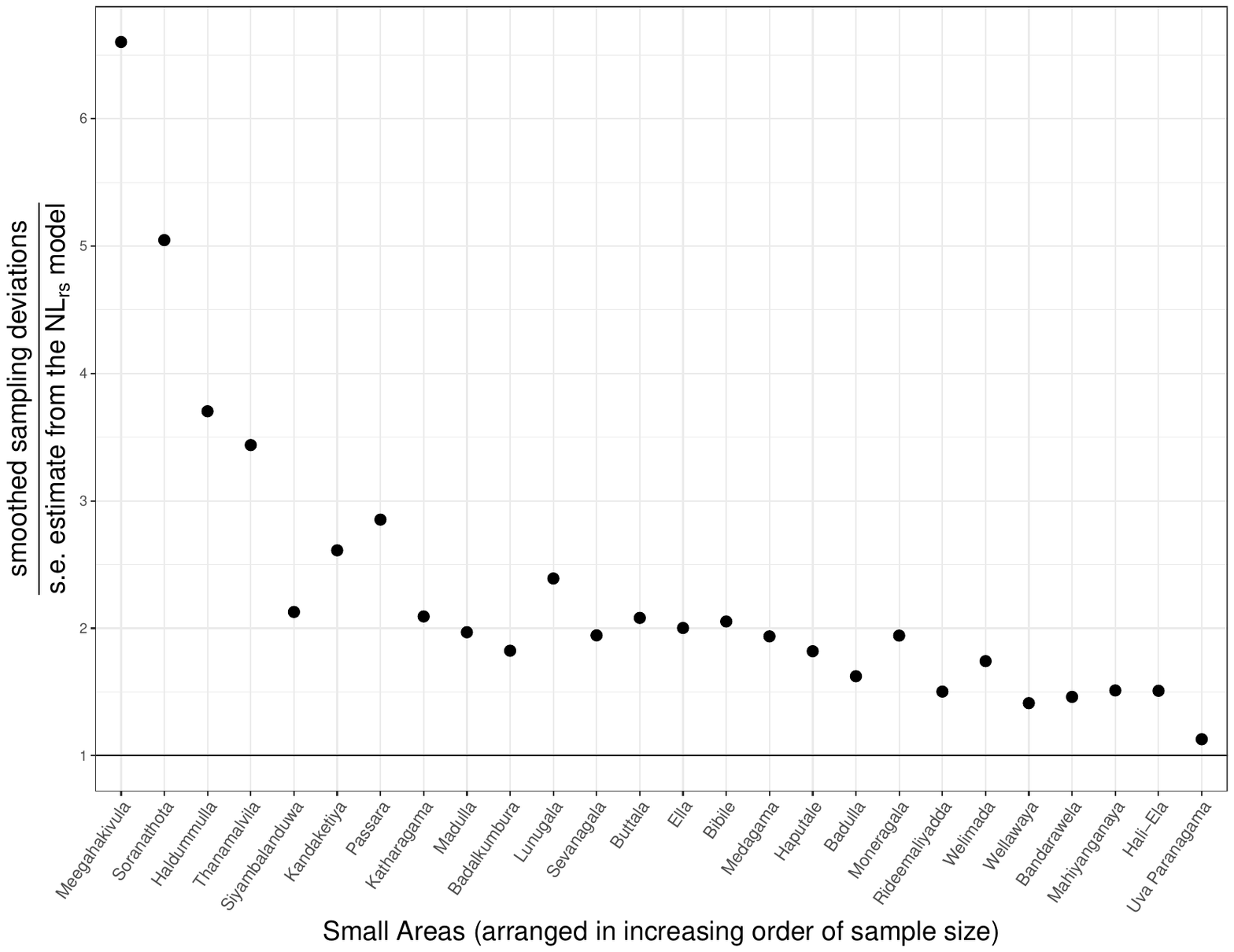}
    \caption{The horizontal line intersects the Y-axis at 1. In X-axis, the small areas are arranged increasing order of sample size. The ratio of the smoothed sampling deviations ($\sqrt{D_i}$) and the standard errors of the posterior small area means from the NL$_{rs}$ model.} 
    \label{fig:ratio_plot}
\end{figure}

As previously discussed, both the Fay-Herriot (FH) model and the Normal-Logistic (NL) model assume known sampling variances of the area-level estimates, modeled at level 1. However, in practice, these variances are not known a priori and must be estimated. Figure \ref{fig:psi_samp} highlights the superiority of the NL$_{rs}$ model, which directly produces posterior standard error estimates. We compare these estimates with the smoothed sampling errors used in fitting the Hierarchical FH and NL models, noting the stability of standard error estimates, particularly in areas with smaller sample sizes. The efficiency gains are more pronounced for such areas, as evident in the ratios depicted in Figure \ref{fig:ratio_plot}, where all ratios exceed 1, with the highest observed for smaller sample sizes.

We also produce district-wise estimates for the two districts within the Uva province. Table \ref{district-direct} gives the district-wise direct survey-based estimates along with their standard errors and 68\% and 95\% confidence intervals. Table \ref{district-rs} gives the district-wise estimates from our Hierarchical Bayes $\text{NL}_{rs}$ model. This table provides the same statistics as the previous one, but along with these, we also report the posterior mean of the benchmarking ratios (benchmarked to the direct district-wise estimates) for the two districts. These numbers are very close to 1. At least at the higher levels (districts are at a higher level than the DSDs--the small areas), model-based estimates are similar to the corresponding survey-weighted estimates, which are known to be reliable at higher levels with the availability of more samples \cite{rao2015small}.

\begin{table}[] 
  \centering 
  \caption{Direct survey based estimates for the two districts with standard errors, 68\% and 95\% Confidence Intervals.} 
  \label{district-direct} 
  
  \begin{adjustbox}{width=0.69\textwidth}
  
\begin{tabular}{@{\extracolsep{5pt}} cccccccc} 
\\[-1.8ex]\hline 
\hline \\[-1.8ex] 
 & district & design based estimate & se & 2.5\% & 16\% & 84\% & 97.5\% \\ 
\hline \\[-1.8ex] 
1 & Badulla & $0.523$ & $0.035$ & $0.453$ & $0.488$ & $0.558$ & $0.593$ \\ 
2 & Moneragala & $0.623$ & $0.042$ & $0.539$ & $0.581$ & $0.665$ & $0.706$ \\
\hline \\[-1.8ex] 
\end{tabular}

\end{adjustbox}

\end{table} 

\begin{table}[] \centering 
  \caption{Estimates for the two districts from the Normal Logistic random sampling model with standard errors, 68\% and 95\% Confidence Intervals. The benchmarking ratio is displayed in the last column.} 
  \label{district-rs} 
  
  \begin{adjustbox}{width=0.69\textwidth}
  
\begin{tabular}{@{\extracolsep{5pt}} ccccccccc} 
\\[-1.8ex]\hline 
\hline \\[-1.8ex] 
 & district & NL$_{rs}$ HB estimate & se & 2.5\% & 16\% & 84\% & 97.5\% & bm\_ratio \\ 
\hline \\[-1.8ex] 
1 & Badulla & $0.529$ & $0.034$ & $0.463$ & $0.495$ & $0.563$ & $0.597$ & $1.011$ \\ 
2 & Moneragala & $0.589$ & $0.039$ & $0.507$ & $0.551$ & $0.628$ & $0.663$ & $0.946$ \\ 
\hline \\[-1.8ex] 

\end{tabular} 

\end{adjustbox}

\end{table} 

Finally, in Table \ref{subset-df}, we display the direct design-based estimates and our NL$_{rs}$ model based estimates for a few interesting small areas.

\begin{table}[] 
\centering 
  \caption{Direct and HB model based estimates along with their standard errors for selected few small areas.} 
 \label{subset-df} 
  
\begin{adjustbox}{width=0.69\textwidth}
\begin{tabular}{@{\extracolsep{5pt}} ccccccccc} 
\\[-1.8ex]\hline 
\hline \\[-1.8ex] 
 & small areas & $n_i$ & $\tilde{n}_i$ & Direct & Direct\_se & NL$_{rs}$ & NL$_{rs}$\_{se} & $\sqrt{D_i}$ \\ 
\hline \\[-1.8ex] 
1 & Meegahakivula & $1$ & $1$ & $0$ & $0$ & $0.619$ & $0.119$ & $0.783$ \\ 
2 & Soranathota & $2$ & $2$ & $0.500$ & $0$ & $0.528$ & $0.110$ & $0.554$ \\ 
3 & Kandaketiya & $11$ & $10.188$ & $0.878$ & $0.015$ & $0.642$ & $0.094$ & $0.245$ \\ 
4 & Lunugala & $23$ & $16.400$ & $0.481$ & $0.246$ & $0.531$ & $0.081$ & $0.193$ \\ 
5 & Sevanagala & $23$ & $20.544$ & $0.725$ & $0.096$ & $0.602$ & $0.089$ & $0.173$ \\ 
6 & Welimada & $45$ & $39.607$ & $0.536$ & $0.092$ & $0.519$ & $0.071$ & $0.124$ \\ 
7 & Uva Paranagama & $73$ & $62.146$ & $0.658$ & $0.120$ & $0.556$ & $0.088$ & $0.099$ \\
\hline \\[-1.8ex] 
\end{tabular} \end{adjustbox}
\end{table}

\subsection{Maps at the lower administrative levels}
In this final subsection of our analyses, we produce maps of different estimates of proportions of poor people by the small areas of the Uva province in Sri Lanka. All of these maps were created using the free and open source GIS software: `QGIS version 2.18.15'. 
\begin{figure}[]
\centering
\begin{subfigure}[t]{.47\linewidth}
  \centering
  \includegraphics[width=0.73\linewidth]{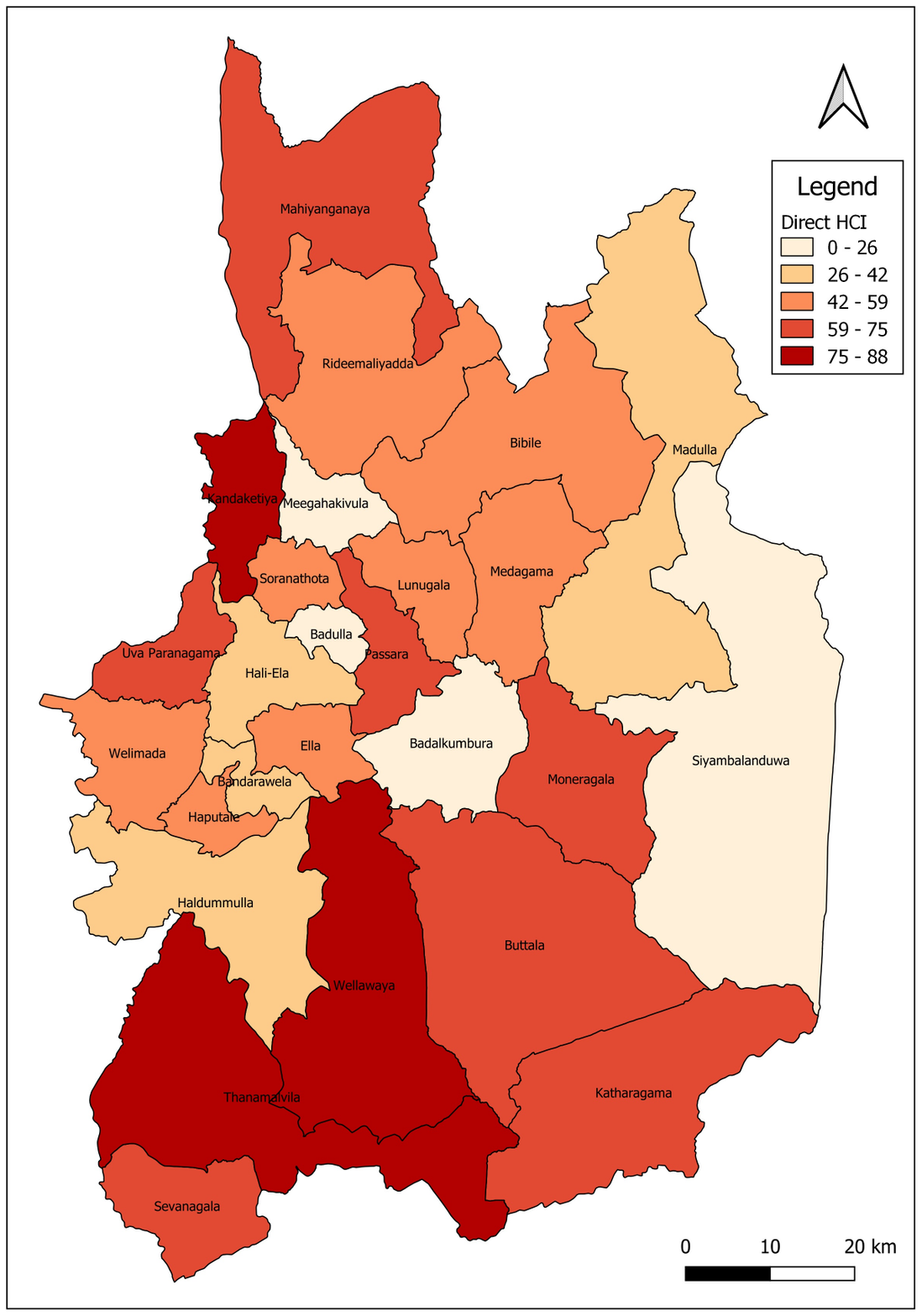}
  \caption{The map of the direct survey based estimates of the percentage of poor people in all the 26 DSDs within the Uva province.}
  \label{fig:direct-map}
\end{subfigure}%
\hfill
\begin{subfigure}[t]{.47\linewidth}
  \centering
  \includegraphics[width=0.73\linewidth]{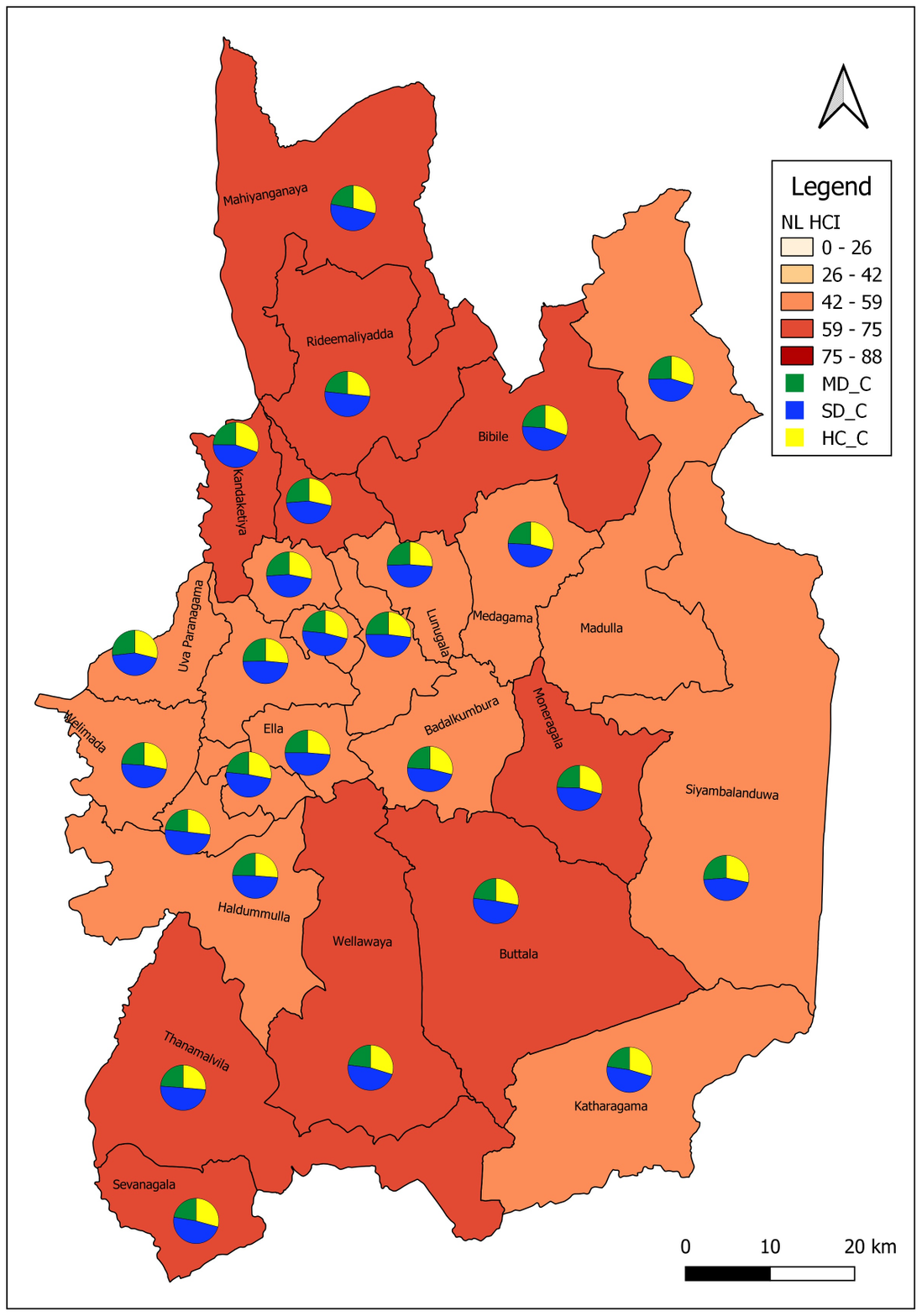}
  \caption{The map of the Hierarchical Bayesian model (NL$_{rs}$) based estimates of the percentages of poor people in all the 26 DSDs within the Uva province. The pie-charts, within each area, report the estimated contribution of the 3 different dimensions towards the composite multidimensional poverty measure.}
  \label{fig:hb-map}
\end{subfigure}
\caption{comparison of Direct, and HB model based estimates}
\end{figure} 

Figure \ref{fig:direct-map} gives the map of the direct survey based estimates of the percentages of poor people for all the 26 small areas, or the Divisional Secretary's Divisions (DSDs) in the Uva province. The problems with the direct survey based estimates for small areas can be observed through this map—highly variable colors (estimates), anywhere from 0\% to $\sim$88\%, especially for the areas with fewer sample sizes. Figure \ref{fig:hb-map} gives the map of the estimates the percentages of poor people for all the 26 small areas from the `best' chosen NL$_{rs}$ model. In contrast to the highly variable direct estimates, we observe much more stabilized estimates, ranging from ~43\% to $\sim$67\%. The pie-charts, placed inside the maps of the DSDs, give the estimated contribution (to the poverty percentages) of the 3 different dimensions that were used to create the composite measure of the multidimensional poverty \cite{deepawansa2022innovative}. `MD\_C' indicates the contribution share of the `Material Deprivation' dimension, `SD\_C' indicates the contribution of the `Social Deprivation' dimension, and finally, `HC\_C' indicates the contribution of the `Human Capital' dimension.
\begin{figure}[]
\centering
\begin{subfigure}[t]{.47\linewidth}
  \centering
  \includegraphics[width=0.73\linewidth]{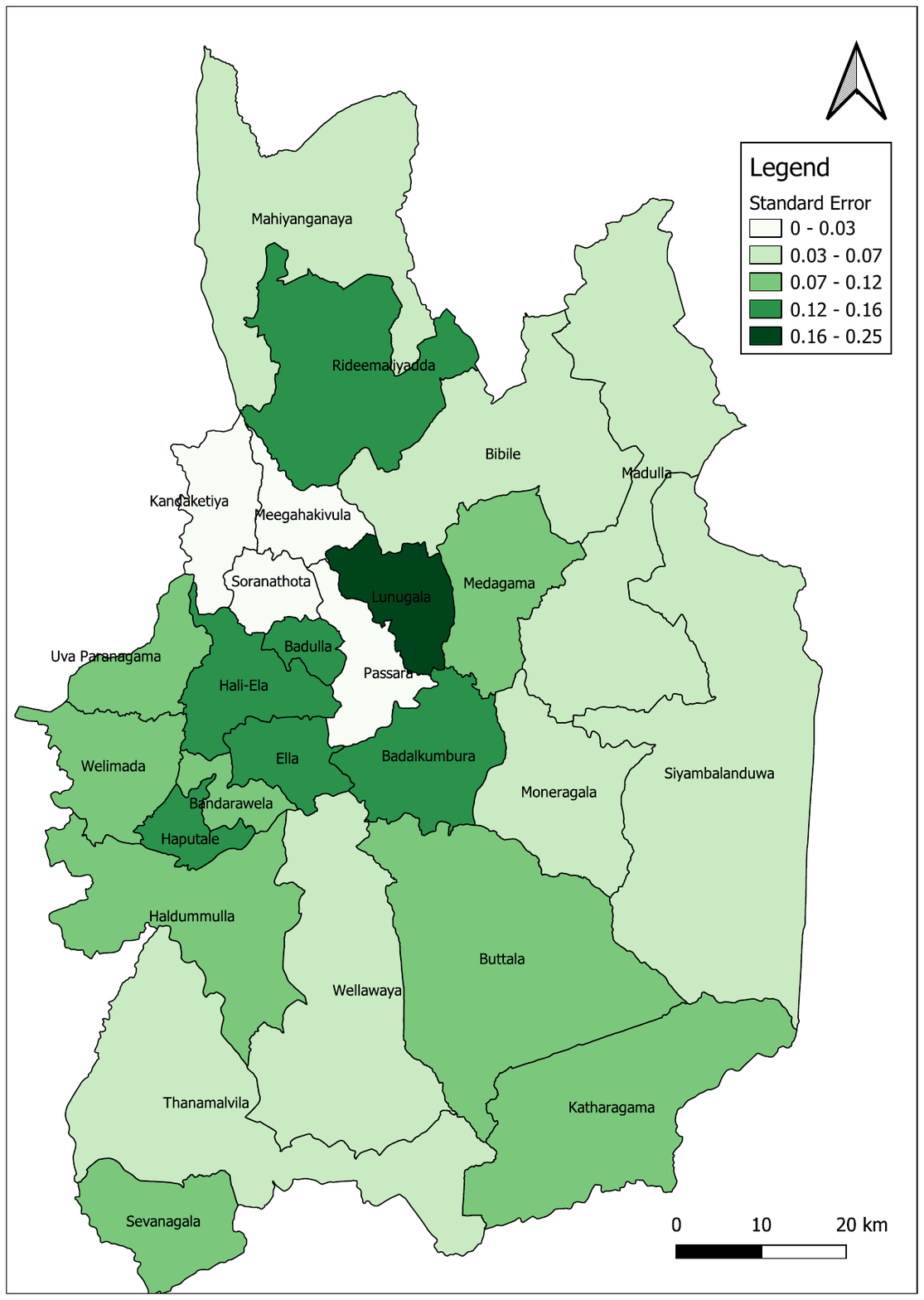}
  \caption{The map of the standard errors of the direct survey based estimates of the proportions of poor people in all the 26 DSDs within the Uva province.}
  \label{fig:direct-se-map}
\end{subfigure}%
\hfill
\begin{subfigure}[t]{.47\linewidth}
  \centering
  \includegraphics[width=0.73\linewidth]{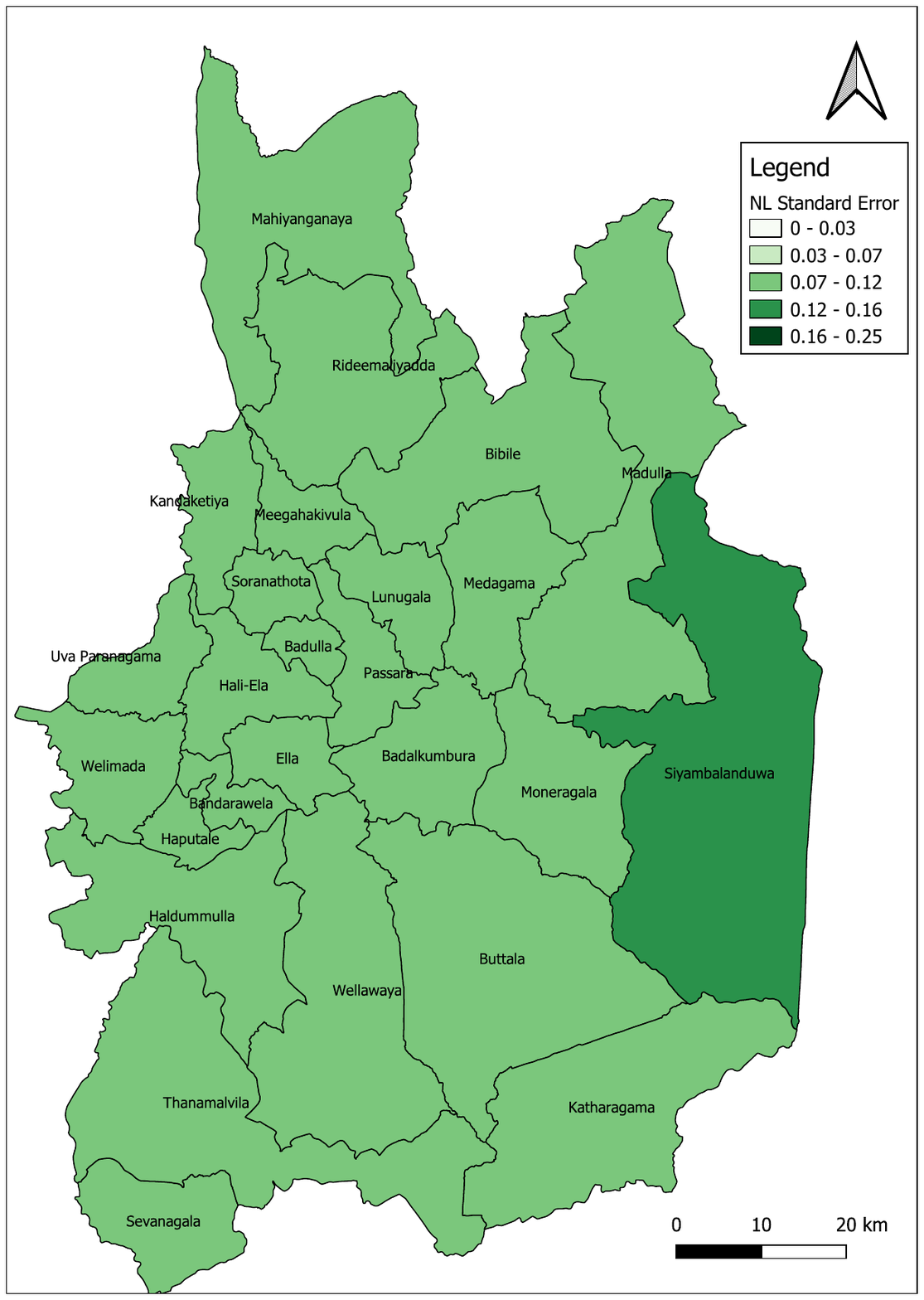}
  \caption{The map of the standard errors of the Hierarchical Bayesian model (NL$_{rs}$) based estimates of the proportions of poor people in all the 26 DSDs within the Uva province.}
  \label{fig:hb-se-map}
\end{subfigure}
\caption{comparison of standard errors of Direct, and HB model based estimates}
\end{figure}
Figure \ref{fig:direct-se-map} gives the map of the standard errors of the direct survey based estimates for all the 26 small areas, or the Divisional Secretary's Divisions (DSDs) of the Uva province. As with the map of the direct estimates, we can also observe highly unstable values of their corresponding standard errors, ranging from 0\% to $\sim$25\%. Figure \ref{fig:hb-se-map} gives the map of the standard errors of HB model (NL$_{rs}$) based estimates of proportions of people who are poor according to the multidimensional measure of poverty for all the 26 DSDs–the small areas of the Uva province in Sri Lanka. We observe much more stabilized values of the standard errors as compared to that of the direct survey based estimates. They range from $\sim$7\% to $\sim$12\%.

\section*{Conclusion}

There is now a sizable literature on developing multidimensional poverty measures. However, the estimation of these measures, especially for small areas, has received little attention. We need both reasonable multidimensional poverty measures and their estimation to make effective evidence-based public policies. In this paper, we address this important research gap by focusing on the estimation of multidimensional poverty measures for small areas.

Using survey data specifically designed for obtaining information on multidimensional poverty, we demonstrate how a Bayesian method can be useful in estimating these poverty rates as well as the relative contributions of different dimensions towards individual-level poverty for small areas. Our study highlights the significance of understanding the contributions of various dimensions to multidimensional poverty, even in areas with minimal economic and infrastructural diversity, such as the Uva province in Sri Lanka. The multidimensional poverty index captures various deprivations experienced simultaneously by the poor, and understanding these contributions helps in identifying policy priorities at the lowest administrative levels, aligning with the SDG goal 1 of ending poverty in all its forms everywhere by 2030.

We considered the synthesis method for multidimensional poverty measures, but our approach to small area estimation can be extended when using counting or fuzzy set methods to define multidimensional poverty. In our multivariate normal hierarchical logistic model, we used a synthetic method for smoothing sampling covariance matrices. For estimating the proportion of impoverished people in small areas, using the multidimensional poverty status, we employed a unique way of determining effective sample sizes, motivated by the complexity of the sampling design of the survey data.

Future research will explore extending the models to incorporate random covariance matrices, further enhancing the accuracy and applicability of multidimensional poverty measures for small areas. Understanding the specific deprivations contributing to poverty in different areas is essential for designing tailored policy interventions, improving living standards, and reducing poverty sustainably.

\section*{Acknowledgements}

We would like to thank Mr. W. B. M. Ranjith Weerasekara, who is a Statistical Officer at the Department of Census and Statistics of Sri Lanka, for his help to create the small area level maps.

\section*{Data Availability}

The unit-level survey data analyzed in this study are not publicly available due to privacy and confidentiality restrictions. However, aggregated summaries and area-level estimates may be available from the corresponding author upon reasonable request and subject to appropriate data use agreements.


\bibliography{ref.bib}

\begin{thebibliography}{}

\bibitem[\protect\citeauthoryear{Alkire and Foster}{Alkire and Foster}{2011}]{alkire2011understandings}
Alkire, S. and J.~Foster (2011).
\newblock Understandings and misunderstandings of multidimensional poverty measurement.
\newblock {\em The Journal of Economic Inequality\/}~{\em 9}, 289--314.

\bibitem[\protect\citeauthoryear{Alkire and Santos}{Alkire and Santos}{2013}]{alkire2013multidimensional}
Alkire, S. and M.~E. Santos (2013).
\newblock A multidimensional approach: Poverty measurement \& beyond.
\newblock {\em Social indicators research\/}~{\em 112\/}(2), 239--257.

\bibitem[\protect\citeauthoryear{Alkire and Seth}{Alkire and Seth}{2015}]{alkire2015multidimensional}
Alkire, S. and S.~Seth (2015).
\newblock Multidimensional poverty reduction in india between 1999 and 2006: Where and how?
\newblock {\em World Development\/}~{\em 72}, 93--108.

\bibitem[\protect\citeauthoryear{Battese, Harter, and Fuller}{Battese et~al.}{1988}]{battese1988error}
Battese, G.~E., R.~M. Harter, and W.~A. Fuller (1988).
\newblock An error-components model for prediction of county crop areas using survey and satellite data.
\newblock {\em Journal of the American Statistical Association\/}~{\em 83\/}(401), 28--36.

\bibitem[\protect\citeauthoryear{Benavent and Morales}{Benavent and Morales}{2016}]{benavent2016multivariate}
Benavent, R. and D.~Morales (2016).
\newblock Multivariate {F}ay--{H}erriot models for small area estimation.
\newblock {\em Computational Statistics \& Data Analysis\/}~{\em 94}, 372--390.

\bibitem[\protect\citeauthoryear{Cerioli and Zani}{Cerioli and Zani}{1990}]{cerioli1990fuzzy}
Cerioli, A. and S.~Zani (1990).
\newblock A fuzzy approach to the measurement of poverty.
\newblock In {\em Income and wealth distribution, inequality and poverty}, pp.\  272--284. Springer.

\bibitem[\protect\citeauthoryear{Das and Lahiri}{Das and Lahiri}{2025}]{das2025approximateHB}
Das, S. and P.~Lahiri (2025).
\newblock Approximate hierarchical {B}ayes small area estimation using {NEF-QVF} and poststratification.
\newblock {\em Survey Methodology\/}.
\newblock In press. Preprint available at arXiv:2210.04980.

\bibitem[\protect\citeauthoryear{Datta, Fay, and Ghosh}{Datta et~al.}{1991}]{datta1991hierarchical}
Datta, G., R.~Fay, and M.~Ghosh (1991).
\newblock Hierarchical and empirical {B}ayes multivariate analysis in small area estimation.
\newblock In {\em Proceedings of Bureau of the Census 1991 Annual Research Conference, US Bureau of the Census, Washington, DC}, pp.\  63--79.

\bibitem[\protect\citeauthoryear{Deepawansa, Dunusinghe, Lahiri, and Gunatilaka}{Deepawansa et~al.}{2022}]{deepawansa2022innovative}
Deepawansa, D., P.~Dunusinghe, P.~Lahiri, and R.~Gunatilaka (2022).
\newblock An innovative approach to measure multidimensional poverty: A synthesis method.
\newblock {\em Statistical Journal of the IAOS\/}~{\em 38\/}(4), 1303--1323.

\bibitem[\protect\citeauthoryear{Dehury and Mohanty}{Dehury and Mohanty}{2015}]{dehury2015regional}
Dehury, B. and S.~K. Mohanty (2015).
\newblock Regional estimates of multidimensional poverty in {I}ndia.
\newblock {\em Economics\/}~{\em 9\/}(1), 20150036.

\bibitem[\protect\citeauthoryear{Erciulescu, Franco, and Lahiri}{Erciulescu et~al.}{2018}]{erciulescu2018chapter}
Erciulescu, A.~L., C.~Franco, and P.~Lahiri (2018).
\newblock Chapter: Use of administrative records in small area estimation.

\bibitem[\protect\citeauthoryear{Fay and Herriot}{Fay and Herriot}{1979}]{fay1979estimates}
Fay, R.~E. and R.~A. Herriot (1979).
\newblock Estimates of income for small places: an application of {J}ames-{S}tein procedures to census data.
\newblock {\em Journal of the American Statistical Association\/}~{\em 74\/}(366a), 269--277.

\bibitem[\protect\citeauthoryear{Gelman, Carlin, Stern, Dunson, Vehtari, and Rubin}{Gelman et~al.}{2013}]{gelman2013bayesian}
Gelman, A., J.~B. Carlin, H.~S. Stern, D.~B. Dunson, A.~Vehtari, and D.~B. Rubin (2013).
\newblock {\em Bayesian data analysis}.
\newblock Chapman and Hall/CRC.

\bibitem[\protect\citeauthoryear{Gelman, Jakulin, Pittau, and Su}{Gelman et~al.}{2008}]{gelman2008weakly}
Gelman, A., A.~Jakulin, M.~G. Pittau, and Y.-S. Su (2008).
\newblock A weakly informative default prior distribution for logistic and other regression models.

\bibitem[\protect\citeauthoryear{Ha, Lahiri, and Parsons}{Ha et~al.}{2014}]{ha2014methods}
Ha, N.~S., P.~Lahiri, and V.~Parsons (2014).
\newblock Methods and results for small area estimation using smoking data from the 2008 national health interview survey.
\newblock {\em Statistics in Medicine\/}~{\em 33\/}(22), 3932--3945.

\bibitem[\protect\citeauthoryear{Kim, Park, and Kim}{Kim et~al.}{2015}]{kim2015small}
Kim, J.~K., S.~Park, and S.-y. Kim (2015).
\newblock Small area estimation combining information from several sources.

\bibitem[\protect\citeauthoryear{Larsen}{Larsen}{2003}]{larsen2003estimation}
Larsen, M.~D. (2003).
\newblock Estimation of small-area proportions using covariates and survey data.
\newblock {\em Journal of Statistical Planning and Inference\/}~{\em 112\/}(1-2), 89--98.

\bibitem[\protect\citeauthoryear{Liu, Lahiri, and Kalton}{Liu et~al.}{2007}]{liu2007hierarchical}
Liu, B., P.~Lahiri, and G.~Kalton (2007).
\newblock Hierarchical {B}ayes modeling of survey-weighted small area proportions.
\newblock In {\em Proceedings of the American Statistical Association, Survey Research Section}, pp.\  3181--3186.

\bibitem[\protect\citeauthoryear{Rao and Molina}{Rao and Molina}{2015}]{rao2015small}
Rao, J.~N. and I.~Molina (2015).
\newblock {\em Small area estimation}.
\newblock John Wiley \& Sons.

\bibitem[\protect\citeauthoryear{Sen}{Sen}{1980}]{sen1980equality}
Sen, A. (1980).
\newblock Equality of what?
\newblock {\em The Tanner Lecture on Human Values\/}.

\bibitem[\protect\citeauthoryear{{Stan Development Team}}{{Stan Development Team}}{2023}]{StanManual}
{Stan Development Team} (2023).
\newblock {\em Stan Modeling Language Users Guide and Reference Manual}.

\bibitem[\protect\citeauthoryear{Sugasawa, Kubokawa, and Rao}{Sugasawa et~al.}{2018}]{sugasawa2018small}
Sugasawa, S., T.~Kubokawa, and J.~Rao (2018).
\newblock Small area estimation via unmatched sampling and linking models.
\newblock {\em Test\/}~{\em 27}, 407--427.

\bibitem[\protect\citeauthoryear{Vehtari, Gelman, and Gabry}{Vehtari et~al.}{2017}]{vehtari2017practical}
Vehtari, A., A.~Gelman, and J.~Gabry (2017).
\newblock Practical {B}ayesian model evaluation using leave-one-out cross-validation and {WAIC}.
\newblock {\em Statistics and computing\/}~{\em 27\/}(5), 1413--1432.

\bibitem[\protect\citeauthoryear{You and Rao}{You and Rao}{2002}]{you2002small}
You, Y. and J.~Rao (2002).
\newblock Small area estimation using unmatched sampling and linking models.
\newblock {\em Canadian Journal of Statistics\/}~{\em 30\/}(1), 3--15.

\end{thebibliography}
\bibliographystyle{chicago}

\end{document}